\documentclass{ieeetran}
\usepackage{macros_arxiv}

\begin{document}

\title{TREET: TRansfer Entropy Estimation via Transformers}

\author{
    \IEEEauthorblockN{
        \uppercase{Omer~Luxembourg}, 
        \uppercase{Dor~Tsur},
        \uppercase{Haim~Permuter},
    }\\
    \IEEEauthorblockA{%
        Ben-Gurion University of the Negev
    }
}

\markboth
{Manuscript submitted to an IEEE journal and currently under review}
{Manuscript submitted to an IEEE journal and currently under review}

\date{}

\maketitle

\begin{abstract}
Transfer entropy (TE) is an information theoretic measure that reveals the directional flow of information between processes, providing valuable insights for a wide range of real-world applications.
This work proposes Transfer Entropy Estimation via Transformers (TREET), a novel attention-based approach for estimating TE for stationary processes. The proposed approach employs Donsker-Varadhan representation to TE and leverages the attention mechanism for the task of neural estimation.
We propose a detailed theoretical and empirical study of the TREET, comparing it to existing methods on a dedicated estimation benchmark.
To increase its applicability, we design an estimated TE optimization scheme that is motivated by the functional representation lemma, and use it to estimate the capacity of communication channels with memory, which is a canonical optimization problem in information theory.
We further demonstrate how an optimized TREET can be used to estimate underlying densities, providing experimental results.
Finally, we apply TREET to feature analysis of patients with Apnea, demonstrating its applicability to real-world physiological data.
Our work, applied with state-of-the-art deep learning methods, opens a new door for communication problems which are yet to be solved.
\end{abstract}

\begin{IEEEkeywords}
Deep Learning,
Information Theory,
Transfer Entropy,
Transformers,
Neural Estimation,
Communication Channels
\end{IEEEkeywords}



\section{Introduction}\label{sec:introduction}
\noindent
\IEEEPARstart{T}{ransfer} entropy (TE), introduced by Schreiber \cite{schreiber2000measuring}, stands as a pivotal information-theoretic measurement that captures the coupling dynamics within temporally evolving systems \cite{nichols2006examining}. Derived as an extension of mutual information (MI), TE is distinctive for its inherent asymmetry, strategically employed in diverse applications for causal analysis \cite{duan2013direct}. TE serves as a robust measure of the directed, asymmetric information flow between two stochastic processes. Specifically, TE quantifies the reduction in uncertainty about a process observations, incorporating the past values of another process to predict the future values of the first \cite{Bossomaier2016AnIT}.

TE has found applications in various domains. In neuroscience, it has proven to be effective in deciphering functional connectivity among and between neurons to various physical tasks \cite{doi:10.1152/jn.01106.2006, ito2011extending, doi:10.1142/S201019451200788X}. Moreover, analysis between visual sensors and movement actuators in embodied cognitive systems can be one via TE \cite{lungarella2006mapping}. 
TE is instrumental in social network analysis, as it quantifies causal relationships between users by measuring the directed flow of information. This enables the detection of influential individuals and the examination of how misinformation propagates through the network \cite{ver2012information}.
\cite{kim2020predicting} utilizes a variant of TE to predict the direction of the US stock market, incorporating TE as input feature. Lately, \cite{bonetti2023causal} suggested a greedy algorithm for feature selection while leveraging the connection between each feature and the target with TE.
Estimating TE is crucial for distinguishing between correlation and causation, as it quantifies the direction and magnitude of information flow between variables, thereby uncovering the underlying causal structures in complex systems

\subsection{Estimation and Optimization of Transfer Entropy}
\noindent
Estimating information theoretic measures such as TE is challenging, especially when the underlying data distribution is unkown \cite{Bossomaier2016AnIT}.
Many methods drows inspiration from MI estimators. Among the estimation methods, Kernel Density Estimation (KDE) \cite{rosenblatt1956remarks}, and $k$-nearest neighbors (KNN) based methods such as Kraskov-Stögbauer-Grassberger (KSG) \cite{kozachenko1987sample, kraskov2004estimating} are noteworthy, despite their struggle with the bias-variance trade-off. 
Granger causality \cite{granger1969investigating}, a statistical concept introduced by Clive W. J. Granger, assesses whether one time series can predict another. Specifically, if the inclusion of past values of a time series $X$ enhances the prediction of another time series $Y$ beyond the information provided by past values of $Y$ alone, then $X$ is said to "Granger-cause" $Y$. This is typically evaluated using vector autoregressive (VAR) models, where the predictive power of past values of $X$ on $Y$ is tested for statistical significance. Notably, Granger causality \cite{granger1969investigating} also serves as a foundational method for TE estimation, particularly in Gaussian joint processes where it is equivalent to TE, highlighting its effectiveness in quantifying information flow between sequences \cite{barnett2012transfer, barnett2009granger}.

Copnet \cite{ma2019estimating}, a copula entropy-based method, builds on empirical density estimation similar to KNN-based approaches and successfully quantifies TE. By establishing the relationship between copula entropy and MI, \cite{ma2019estimating} proposed a formulation of TE using four copula entropy terms.

Recent advancements have embraced neural estimation approaches like the Donsker-Varadhan (DV) and Nguyen-Nowozin-Jordan variational formulas for Kullback-Leibler (KL) and $f$-divergence to accurately estimate MI, directed information (DI) rate, and TE \cite{donsker1983asymptotic, belghazi2018mutual, tsur2023neural, zhang2019itene}. These methods include MI neural estimation (MINE) for MI, DI neural estimation (DINE) for DI rate, and TE neural estimation (TENE) for TE, solving the optimization problems posed by variational formulas using NNs.

TENE \cite{zhang2019itene} uses the classifier-based conditional MI estimator \cite{mukherjee2020ccmi}, which estimates two MI terms that are subtracted to obtain TE. Besides using NNs for optimization instead of a non-parametric method, TENE reduce the number of terms in the TE formula to two MI values, unlike Copnet. However, it suffers from input dimensionality issues when estimating higher-order TE, since increasing the sequence length to estimate larger-order TE introduces input dimensionality challenges in the DV-based variational formulation.
In contrast, \cite{tsur2023neural} estimate DI rate, a closely related measure to TE, using RNNs, enabling estimation for theoretically infinite memory sequences. Nevertheless, DINE is designed to estimate information flow over infinite sequences, which corresponds to the DI rate and is not well-suited for finite-order TE estimation.

NNs have revolutionized fields such as computer vision and natural language processing, with transformers now dominating time-series analysis, overtaking RNNs due to their superior performance \cite{vaswani2017attention, zhou2021informer, wu2021autoformer, zhou2022fedformer}. Despite the widespread adoption of MINE for information-theoretic estimation, it has been subject to scrutiny due to its limitations, such as high variance in gradient estimates, bias in MI estimation, and training instability under high-dimensional settings \cite{song2019understanding, mcallester2020formal, choi2022combating}.

To address these challenges, we introduce TREET, a transformer-based neural estimator of transfer entropy for high-dimensional continuous data.

\subsection{Contributions}
\raggedbottom
In this work, we introduce TREET, a novel TE estimator that leaverages the attnetion mechanism. Our method is grounded in the DV representation, leveraging attention-based NNs adapted to meet the structural constraints of DV optimization. Theoretically, we establish the consistency of TREET and develop an auxiliary neural distribution generator (NDG), an input distribution generative module to facilitate TE optimization, utilizing transformers.
Empirically, we showcase the versatility of TREET across various applications. 
\begin{itemize}
    \item Show that TREET outperforms TENE and Copnet by extending the benchmark presented in \cite{zhang2019itene} to address longer order TE, particulary in scenarios of high valued TE and long temporal contexts.
    \item Demonstrate the joint optimization of TREET and the NDG for the estimation of the capacities of channels with memory, supported by theoretical validations. Experiments on long-memory channels emphasize TREET's ability to capture and utilize extensive historical dependencies while accurate estimating their channel capacities. 
    \item Derive a density estimator of the underlying process, which emerges as a byproduct of the TREET optimization.
    \item Apply TREET to the Apnea dataset \cite{rigney1993apnea, goldberger2000apnea} for feature analysis, illustrating its utility in uncovering causal relationships within real-world data, paving the way for broader applications in future research.
\end{itemize}

\subsection{Organizations}
\raggedbottom
The reminder of the paper is organized as follows, Section \ref{sec:background} provides background on information theory and NNs. Section \ref{sec:est_te} presents the TREET, its theoretical guarantees and practical implementations, whereas the optimization of  TREET is described in Section \ref{sec:optimization_te}. Experimental results on TE dedicated benchmark, channel capacity estimation and it's memory analysis, probability density estimation and features analysis on real-world data are shown in Section \ref{sec:results}. Section \ref{sec:conclusions_future_work} concludes the paper and discuss future research potentials.


\section{Background}\label{sec:background}
\raggedbottom
In this section, we elaborate on the preliminaries necessary to present our method. We familiarize the reader with our notation and provide the formal definition of TE, then relate it to DI. Subsequently, we introduce the concept of transformer NN, define them, and discuss the theorem of universal approximation. Lastly, we present the use of NN as estimators for information theoretic measures.

\subsection{Notation}
\raggedbottom
Calligraphic letters, such as $\cX$, denote subsets of the $d$-dimensional Euclidean space, $\RR^d$. Expectations are represented by $\EE$, with all random variables defined in the probability space $(\Omega, \cF, \PP)$. The collection of Borel probability measures on $\cX$ is indicated as $\cP(\cX)$, and $\lebmeas(\cX)$ specifically refers to those measures that are absolutely continuous with respect to Lebesgue measure, with their densities denoted by lowercase $p$. Random variables and vectors are uppercase, e.g., $X$, and stochastic processes are in blackboard bold, e.g., $\XX:=(X_t)_{t\in\ZZ}$ for discrete time $t$.

The sequence of $l$ samples from time $t$ in process $\XX$ is $X_t^{t+l}:=[X_t,\ldots,X_{t+l}]^\top$, and for stationary processes, $X^l:=[X_0,X_1,\ldots,X_l]^\top$. For $\cQ\in\lebmeas(\cX)$ with PDF $q$, the cross entropy between $P$ and $Q$ is $\sh_{\mathsf{CE}}(P,Q):=-\EE_P[\log q]$. Differential entropy of $X\sim P$ is $\sh(X):=\sh_{\mathsf{CE}}(P,P)$. If $Q\ll P$, the KL divergence $\DKL(P\|Q):=\EE_P[\log\frac{\mathrm{d}P}{\mathrm{d}Q}]$. MI between $(X,Y)\sim P_{XY}$ is $\sI(X;Y) := \DKL(P_{XY}\|P_X\otimes P_Y)$. Conditional KL divergence for $P_{Y|X}, Q_{Y|X}$ given $X\sim P_X$ is $\DKL(P_{Y|X} \| Q_{Y|X} | P_X)$, and conditional MI for $(X,Y,Z)\sim P_{XYZ}$ is $\sI(X;Y|Z):=\DKL (P_{XY|Z}\|P_{X|Z}\otimes P_{Y|Z}|P_Z)$.

For a comprehensive list of all symbols and their definitions, please refer to Appendix~\ref{sec:appendix_symbols}.

\subsection{Transfer Entropy}
\raggedbottom
TE quantifies causal influence from the past of one sequence on the present of another, formally given as follows

\begin{definition}[Transfer Entropy]\label{def:transfer_entropy}
    For jointly distibuted processes $\XX$ and $\YY$ and $k,l<\infty$, the TE, with parameters $(k,l)$, is given by
    \begin{equation}\label{eq:te_with2parms}
        \sTE(t;k,l):=\sI\left(X^t_{t-l};Y_t | Y^{t-1}_{t-k}\right).
    \end{equation}
\end{definition}
When the considered processes are jointly stationary, the temporal index $t$ in \eqref{eq:te_with2parms} can be omitted, and TE is written as
\begin{equation}\label{eq:te_with2parms_stationary}
    \sTE(k,l):=\sI\left(X^l;Y_l | Y^{l-1}_{l-k}\right).
\end{equation}
Note that our definition of TE includes $X_t$, whereas other definitions \cite{schreiber2000measuring, liu2012relationship, barnett2012transfer, zhang2019itene} 
define TE as $\sI(X^{l-1};Y_l|Y^{l-1}_{l-k})$. 
The main difference introduced by this change is that the definition reduces to MI when the processes are jointly independent and identically distributed (i.i.d.), i.e., $\sTE(0,0)=I(X;Y)$. When $k=l$, the abbreviated notation $\sTE(l)$ is used.

\subsection{Relation to Directed Information}\label{sec:te_and_di_relations}
\raggedbottom
DI \cite{massey1990causality,kramer1998directed} is given by
\begin{align*}\label{eq:di}
    \sI\left(X^n \rightarrow{} Y^n \right) &:= \sum_{i=0}^{n-1} \sI\left( X^i ; Y_i | Y^{i-1}\right)\numberthis \\
    \sI\left(X^n \rightarrow{} Y^n \right) &= \sum_{i=0}^{n-1} \sTE(i).\numberthis
\end{align*}
When $(X^n,Y^n)$ are the $n$-fold projections of stochastic processes $\XX$ and $\YY$, which \eqref{eq:te_with2parms_stationary} shows that DI is the sum of TE terms. The DI \textit{rate} can be defined as

\begin{align*}
    \sI(\XX\to\YY) &:= \lim_{n \to \infty} \frac{1}{n}\sI\left( X^n \rightarrow{} Y^n\right) \\
    &\stackrel{(a)}{=} \lim_{n\to\infty} \sI(X^n;Y_n|Y^{n-1})\numberthis \label{eq:di_rate},
\end{align*}
and $(a)$ holds due to stationarity \cite{tsur2023neural}.
The DI rate can be therefore observed as a limit case of TE, where the limit of TE exists since $\sTE(l)=\sh(Y_l|Y^{l-1}) - \sh(Y_l|X^l,Y^{l-1})$ and each conditional entropy is non-negative, decreasing for increasing $l$ (conditioning reduces entropy).
Moreover, under appropriate Markov assumptions, $Y_l - Y^{l-1} - Y^{-1}_{-\infty}, Y_l - (X^l,Y^{l-1})-(X^{-1}_{-\infty},Y^{-1}_{-\infty})$, an equality between TE and DI can be established,
\begin{lemma}\label{lemma:te_di_rate_relation_markov}
    Define markov property $Z_n - Z_{n-1} - Z^{n-2}$ as $P_{Z_n | Z^{n-1}}=P_{Z_n|Z_{n-1}}$.
    Let $\XX$ and $\YY$ be two jointly stationary processes such that the following markov property holds,
        \begin{align*}
        &Y_l - Y^{l-1} - Y^{-1}_{-\infty},\\
        &Y_l - (X^l,Y^{l-1})-(X^{-1}_{-\infty},Y^{-1}_{-\infty}),
    \end{align*}
    for $l\in\NN$. Then, for $m\in\NN, m \geq l$,
    \begin{equation}\label{eq:tem_tel}
        \sTE(m)=\sTE(l),
    \end{equation}
    and
    \begin{equation}\label{eq:te_di_rate_relation_markov}
        \sTE(m)=\sI\left( \XX \to \YY \right).
    \end{equation}
\end{lemma}
Refer to Appendix \ref{sec:proof_lemma_te_di_rate_relation_markov} for detailed proof.

\subsection{Attention Mechanism and Transformers}\label{sec:neural_networks_and_transformers}
\raggedbottom
NNs capabilities can be enhanced by the attention\footnote{This paper considers the dot-product attention.} mechanism \cite{vaswani2017attention}, which selects various combinations of inputs according to their significance to the predictions. The attention can address time series datasets, while weighting over each time input. 
Attention comprises \textit{queries}, which represent the current temporal focus, \textit{keys}, which match against the queries to determine relevance, and \textit{values}, that contain the NN inputs to be weighted and act as a memory function in time-series.

\begin{definition}[Attention]\label{def:attention_dotproduct}
    Let $W_Q,W_K,W_V\in\RR^{x \times d}$ and let $Q=XW_Q,K=XW_K$ and $V=XW_V$ be the \textit{queries}, \textit{keys} and \textit{values}, $X=(x_1,x_2,\ldots,x_n)$, $\forall i: x_i\in\RR^{d_x}$. Then the \textit{attention} is given by,
    \[
    \attn(X) = \mathsf{softmax}\left(Q K^\top \right) V,
    \]
    where $\mathsf{softmax}(Z)_j={\exp{[Z_j]}}/{\sum_{m=1}^{n} \exp{[Z_m]}}$ where $Z\in\RR^{n}$, is performed for each column of the dot-product between queries and keys separately.
\end{definition}

Transformers utilize positional encoding (PE), which maps each input to the attention according to its ordinal index, and is applied before the attention for time series applications to deformation of the sequence structure. 
Attention can be generalized by considering several heads which operate in parallel. Multi-head attention is given by:
\begin{equation}
    \mathsf{MultiHead}\text{-}\attn(X)=[H_1,\ldots,H_h]W_0,
\end{equation}
where
\begin{equation}
    H_i = \mathsf{softmax} \left (Q^{(i)} {K^{(i)}}^\top \right) V^{(i)}, \quad i=1,\ldots,n,
\end{equation}
$W_0 \in \RR^{dh \times d}$ is learnable parameter, and $Q^{(i)}, K^{(i)}, V^{(i)}$ are the $i^{\text{th}}$ learnable projections of queries keys and values \cite{vaswani2017attention}. A transformer block is a sequence-to-sequence function, which maps input sequence to an output sequence. Each transformer block consists of attention layer and time-wise feed-forward layer.
Formally, a transformer function class is given as follows \cite{yun2019transformers}
\begin{definition}[Transformer function class]\label{def:transformer_func_class} 
    Let $d_i,d_o,l,v\in\NN$. The class of transformers with $v$ neurons, denoted $\GTF^{(d_i, d_o, l, v)}: \RR^{d_i\times l}\to\RR^{d_o \times l}$, is the set of discrete-time with the following structure:
    \begin{subequations}
        \begin{align*}
            &\Xpe = W_1 \cdot X + E, \numberthis\\
            &\attn(\Xpe) = \Xpe +\sum_{i=1}^h W_O^iW_V^i \Xpe \\
            & \qquad \qquad \cdot \mathsf{softmax} \left[ (W_K^i \Xpe)^\top (W_Q^i\Xpe)\right],\numberthis \label{eq:attention_tfc}\\
            &Y=\ff(\Xpe) = \attn(\Xpe) + W_3 \cdot \sigma_\sR \\
            & \qquad \qquad \left(W_2 \cdot \attn(\Xpe) + b_1 \mathbf{1}_l^\top \right) + b_2 \mathbf{1}_l^\top,\numberthis
        \end{align*}
    \end{subequations}
    where $X\in \RR^{d_i \times l}$ is the input sequence of $l$ samples, $Y\in\RR^{d_o \times l}$ is the transformer output, $W_1 \in \RR^{d_e \times d_i}, W_O^i \in \RR^{d_e \times d_m}, W_Q^i, W_K^i, W_V^i \in \RR^{d_m \times d_e}, W_2 \in \RR^{d_r \times d_e}, W_3 \in \RR^{d_e \times d_r}, b_1\in \RR^{d_r}, b_2 \in \RR^{d_e}$ are the weights and biases of the network, $E \in \RR^{d_e \times l}$ is the PE of the input.
    The number of heads $h$ and the head size $d_m$ are the parameters of the multi-head attention and $d_r$ is the hidden dimension of the feed-forward (FF) layer.
    Assuming that the input is an additive product with it's positional encoding product, before the transformer blocks.
    The class of transformers with dimensions $(d_i, d_o, l)$ is thus given by
    \begin{equation}\label{eq:transformer_func_class_set}
        \GTF^{(d_i,d_o, l)} := \bigcup_{v\in\NN} \GTF^{(d_i, d_o, l, v)}.
    \end{equation}
\end{definition}
Transformers are a universal approximation class of sequence to sequence mappings \cite{yun2019transformers}:
\begin{theorem}[Universal approx. for transformers]\label{theorem:ua_transformers}
    Let $\epsilon>0$, $l \in \NN$. $\cU\subset \RR^{T\times d_i}, \cZ\subset \RR^{l\times d_o}$ be open sets, and $f:\cU \to \cZ$ be a continuous vector-valued function. Then, there exist $v \in \NN$ and a $v$-neuron transformer $g\in\GTF^{(d_i,d_o,l,v)}$ (as in Definition \ref{def:transformer_func_class}, such that for any sequence of inputs $\{u^l\} \in \cU$ and sequence of outputs $\{z^l\} \in \cZ$, we have
    \begin{equation}
        \left\| f(u^l) - g(u^l) \right\|_1 \leq \epsilon,
    \end{equation}
\end{theorem}

Theorem \ref{theorem:ua_transformers} establishes the universal approximation capabilities of transformers for sequence-to-sequence mappings, whereas many real-world applications, such as time-series forecasting and information flow estimation, require models that respect the inherent causal structure of sequential data. To address this, the class of \textit{causal transformers} is utilized, $\GTFC$, which is built upon $\GTF$ and enforces a strict temporal dependency, ensuring that each output at certain time step is computed only from past and present inputs. This causal structure is crucial for tasks such as TE estimation, where information flow must be inferred without accessing future observations.
To this end, the notation of causal functions is used
\begin{definition}[Causal Function]\label{def:causal_function}
    Let $\mathcal{F}: \mathbb{R}^{d_u \times L} \to \mathbb{R}^{d_z\times L}$ be a function for $d_u,d_z,L \in \NN$. For a series of inputs $U=\{u_t\}_{t=1}^L$ and function outputs $Z=\{z_t\}_{t=1}^L$, the function $\mathcal{F}$ is as a causal function if, for any $t_0$ the output $z_{t_0}$ depends only on the inputs $\{u_t\}_{t\leq t_0}$.
\end{definition}
To enforce causality in the mapping function, a \textit{causal mask} is applied to the attention scores \cite{vaswani2017attention}, ensuring that each output only depends on past and present inputs, as defined in Definition \ref{def:causal_function}. Notably, only the attention mechanism performs time mixing, so applying the causal mask does not alter the rest of the transformer architecture. The causal mask, denoted as $M\in\RR^{l \times l}$, is applied element-wise to the dot-product of keys and queries before the softmax operation and is given by
\begin{equation}
    M_{[i,j]} = 
    \begin{cases}
        1 & \text{if} j \leq i \\
        -\infty & \text{otherwise}
    \end{cases}     
\end{equation}
where $-\infty$ nullifies the corresponding entries after the softmax operation. 
Hence, \eqref{eq:attention_tfc} is written as,
\begin{align*}
    &\attn(\Xpe) = \Xpe +\sum_{i=1}^h W_O^iW_V^i \Xpe \\
            & \qquad \cdot \mathsf{softmax} \left[ (W_K^i \Xpe)^\top (W_Q^i\Xpe) \odot M \right],\numberthis \label{eq:attention_ctfc}
\end{align*}
where $\odot$ is the Hadamard (element-wise) product, meaning that the multiplication is performed entry-wise between matrices of the same dimensions.
Although changing the dot-product operation of the attention, the model remains consistent with the universality framework, which is grounded in the architecture's capacity for pair-wise operations rather than being constrained by the specifics of the attention mechanism \cite{yun2019transformers}.

\subsection{Neural Estimation}
\noindent
Neural estimation leverages NNs to approximate and optimize statistical functionals, such as mutual information and statistical divergences.
Neural estimators often utilize a variational formula, such as the DV representation \cite[Theorem 3.2]{donsker1983asymptotic}.
\begin{theorem}[DV representation]\label{theorem:dv_representation}
    For any, $P, Q \in \cP(\cX)$, we have
    \begin{equation}\label{eq:dv_represnetation}
        \DKL(P\|Q)=\sup_{f:\cX\rightarrow{}\RR} \EE_P [f] -\log \left(\EE_Q [e^f]\right),
    \end{equation}
    where the supremum is taken over all measurable functions $f$ with finite expectations.
\end{theorem}
To obtain an estimate from the DV representation, the class of functions is approximated with the class of NNs and expectations are replaced with sample means \cite{belghazi2018mutual, molavipour2021neural, zhang2019itene, tsur2023neural}.
A provably consistent neural estimator of TE is developed, that utilizes the power of the attention mechanism.


\section{Estimation of Transfer Entropy}\label{sec:est_te}
\noindent
Transformers excel at capturing long-range dependencies in time series, offering superior scalability and training stability compared to RNNs \cite{vaswani2017attention}. While DINE \cite{tsur2023neural}, built on RNNs, helped mitigate the high variance and bias of neural information-theoretic estimators, self-attention’s global receptive field and parallelism allow it to further improve neural estimation in high-dimensional settings.

In this paper we harness to computational power of transformers architecture and modify the attention mechanism to result with TREET, a new estimator of TE for high dimensional continuous data.
We begin by deriving the estimator, then account for its theoretical guarantees.
Finally, implementation details are provided, outlining the modifications of attention to neural estimation.

\subsection{Estimator Derivation}\label{subsec:estimator}
\noindent
The TREET provides an estimate of the TE \eqref{eq:te_with2parms} from a set of samples $D_{n,l}=(X^{n+l}, Y^{n+l})\sim P_{X^{n+l}, Y^{n+l}}$, where $l$ is the memory order of the TE. 
Assuming that $l\ll n$ and thus omitting the dependence on $l$ in the dataset notation, i.e. $D_{n}:=D_{n,l}$.
To derive TREET, first we represent TE as subtraction of KL divergences, w.r.t. some absolutely continuous reference distribution $\tilde{P}_Y$ over the alphabet $\cY$\footnote{If $\cY$ is not bounded, the maximal bounds can be set, regarding to the dataset properties.}.
We propose the following.
\begin{lemma}[TE as KL Divergences]\label{lemma:te_dkl}
    TE decomposes as
    \begin{equation}\label{eq:te_dkl}
        \sTE(l) = \DYXT - \DYT,
    \end{equation}
    where
    \begin{subequations}\label{eq:dy_dyxt_def}
        \begin{align*}
            &\DYT := \DKL \left(P_{Y_l|Y^{l-1}} \| \tilde{P}_{Y} \Big{|} P_{Y^{l-1}} \right),\numberthis\\
            &\DYXT := \DKL \left(P_{Y_l|Y^{l-1}X^l} \| \tilde{P}_{Y} \Big{|} P_{Y^{l-1}X^{l}} \right), \numberthis
        \end{align*}
    \end{subequations}
    and the conditional KL divergence is $\DKL(P_{X|Z}\|P_{Y|Z}|P_Z):=\EE_Z[\DKL(P_{X|Z}\|P_{Y|Z})]$.
\end{lemma}  
Lemma \ref{lemma:te_dkl} is proved in Section \ref{subsec:te_dkl_proof}, and follows basic information theoretic properties of TE and KL divergences.
Let $\GTFC^Y:=\GTFC^{(d_y,1,l,v_y)}, \GTFC^{XY}:=\GTFC^{(d_x+d_y,1,l,v_{xy})}$ be sets of causal transformer architectures for $l,d_y,d_x,v_y,v_{xy} \in \NN$.
Each KL term can be approximated using the DV representation \eqref{eq:dv_represnetation} as follows:
\begin{subequations}\label{eq:dyt_and_dyxt_expectation}
   \begin{align*}
        \DYT =& \sup_{g_y \in \GTFC^Y} \EE \left[ g_y\left(Y^{l}\right)\right] \\
        &  - \log \left( \EE \left[ e^{g_y \left(\tilde{Y}_l, Y^{l-1}\right)}\right] \right),\label{eq:dyt_expectation}\numberthis\\
        \DYXT =& \sup_{g_{xy} \in \GTFC^{XY}} \EE \left[ g_{xy}\left(Y^{l},X^{l}\right)\right] \\
        &  - \log \left( \EE \left[ e^{g_{xy}\left(\tilde{Y}_l, Y^{l-1}, X^{l}\right)}\right] \right).\numberthis
    \end{align*} 
\end{subequations}
where $g_y\in\GTFC^{Y}, g_{xy}\in\GTFC^{XY}$.
Finally, replacing expectations with sample means in \eqref{eq:dyt_and_dyxt_expectation}, yields the TREET, given by
\begin{subequations}\label{eq:tene_objective_estimator}
    \begin{align*}
        &\tene(\Dn;l)\\
        &=\sup_{g_{xy}\in \GTFC ^{XY}} \hDYXT(\Dn, g_{xy}) \\
        &\quad  -\sup_{g_{y}\in \GTFC ^{Y}} \hDYT(\Dn, g_{y}),\label{eq:tene_estimator}\numberthis
    \end{align*}
\end{subequations}
where
\begingroup
\allowdisplaybreaks
\begin{subequations}\label{eq:dyt_and_dyxt}
    \begin{align*}
        &\hDYT(\Dn, g_y)
        := \frac{1}{n}\sum_{i=1}^{n} g_y\left(
            Y_{i}^{i+l}
        \right)& \\
        & \quad \qquad \qquad \qquad -\log\left(
            \frac{1}{n} \sum_{i=1}^{n} e^{g_y
            \left( 
                \tilde{Y}_{i+l}, Y_{i}^{i+l-1}
            \right)}
        \right),& \numberthis \label{eq:dyt_and_dyxt_a}\\
        &\hDYXT(\Dn, g_{xy})
        := \frac{1}{n}\sum_{i=1}^{n}g_{xy}
        \left(
            Y_{i}^{i+l}, X_{i}^{i+l}
        \right)\\
        & \quad \qquad \qquad -\log
        \left(
            \frac{1}{n} \sum_{i=1}^{n} e^{g_{xy}
            \left(
                \tilde{Y}_{i+l}, Y_{i}^{i+l-1}, X_{i}^{i+l}
            \right)}
        \right).\numberthis \label{eq:dyt_and_dyxt_b}
    \end{align*}
\end{subequations}
\endgroup
where $\tilde{Y}$ is i.i.d. under absolutely continuous reference measure under the alphabet $\cY$, and 
\begin{align*}
    &\sup_{g_{xy}\in \GTFC ^{XY}} \hDYXT(\Dn, g_{xy})=\DYXT, \\
    &\sup_{g_{y}\in \GTFC ^{Y}} \hDYT(\Dn, g_{y})=\DYT.
\end{align*}
The TREET \eqref{eq:tene_objective_estimator} is consequently given as the subtraction of the solutions of two optimization problems \eqref{eq:dyt_and_dyxt_a}, \eqref{eq:dyt_and_dyxt_b}, each optimizing its own NN. The optimization is performed via mini-batch gradient ascent with the corresponding model $g_y, g_{xy}$ and the dataset $\Dn$.
Next, we prove that the TREET implemented with causal transformers is a consistent estimator of TE. 

\begin{figure}[t]
    \centering
    \hspace{-3em}
    \scalebox{0.55}{%

\usetikzlibrary{arrows.meta, chains, positioning, calc}

\tikzstyle{block_big} = [rectangle, rounded corners, minimum width=3cm, minimum height=3cm, align=center, draw=black, fill=black!5]
\tikzstyle{block_small} = [rectangle, rounded corners, minimum width=1.5cm, minimum height=1.5cm, align=center, draw=black, fill=black!5]

\tikzstyle{block_D} = [rectangle, rounded corners, minimum width=2.5cm, minimum height=1.4cm, align=center, draw=black, fill=black!5]
\tikzstyle{block_E} = [rectangle, rounded corners, minimum width=0.8cm, minimum height=0.8cm, align=center, draw=black, fill=black!5]
\tikzstyle{block_C} = [rectangle, minimum width=1cm, minimum height=0.01cm, draw=white, fill=white]

\tikzstyle{invisible} = [rectangle]


\begin{tikzpicture}


\node (input_yi) at (-1,0) {};


\node (input_yi_2) at (-1,-1.7) {};

\node [block_small, right=0.0cm of input_yi_2] (reference) {Reference\\Sampler};


\definecolor{transformer_color}{RGB}{232,232,247}
\node [block_big, fill=transformer_color] (transformer) at (4,-0.85) {\large TREET\\ \large $\theta_y$};

\node [block_big] (loss) at (10.5, -0.85) {\large{DV Loss}\\ $\hDYT(\Dn, g_{\theta_y})$ 
};

\draw [arrows = {-Latex[width=7pt, length=7pt]}] (reference) -- node[below=1mm] {$\tilde{Y}_t$} ([yshift=-0.85cm] transformer.west);

\draw [arrows = {-Latex[width=7pt, length=7pt]}] (input_yi) -- node[above right=1mm and 5.5mm] {$Y^t_{t-l}$} ([yshift=0.85cm] transformer.west);

\draw [-{Implies},double,-{Latex[width=3mm]}] ([yshift=0.85cm]transformer.east) -- node[above=1mm] { \large$g_{\theta_y}(Y_t,Y_{t-l}^{t-1})$} node[below=1mm] { \large$g_{\theta_y}(\tilde{Y}_t,Y_{t-l}^{t-1})$} ([yshift=0.85cm]loss.west);

\draw [arrows = {-Latex[width=7pt, length=7pt]}, dashed] ([yshift=-0.85cm]loss.west) -- node[below=1mm] {\large $\nabla_{\theta_y}\widehat{\mathsf{D}}_Y(D_n,g_{\theta_y})$} ([yshift=-0.85cm]transformer.east);

\end{tikzpicture}

%
    }
    \caption{The estimator architecture for the calculation of $\hDYT(\Dn, g_{\theta_{y}})$.}
    \label{fig:tene_dy}
\end{figure}

\subsection{Theoretical Guarantees}\label{subsec:theoretical_guarantees}
\noindent
The TREET consistency is established when the joint process $(\XX, \YY)$ is stationary and the TREET networks are implemented with the class of causal transformers. We have the following:
\begin{theorem}[TREET consistency]\label{theorem:tene_consistency}
    Let $\XX$ and $\YY$ be jointly stationary, ergodic stochastic processes. The TREET is a strongly consistent estimator of $\sTE(l)$ for $l\in\NN$, i.e. $\PP-a.s.$ for every $\epsilon>0$ there exists and $N \in \NN$ such that for every $n>N$ we have
    \begin{equation}\label{eq:tene_consistency}
        \left| \tene(\Dn;l) - \sTE(l) \right| \leq \epsilon
    \end{equation}
    where $l$ is the memory parameter of the TE.
\end{theorem}
The proof following the steps of 1) representation step - represents TE as a subtraction of two DV potentials, 
2) estimation step - proves that the DV potentials is achievable by empirical mean of a given set of samples, 
and 3) approximation step - shows that the estimator built upon causal transformers converges to TE with the corresponding memory parameter. The proof is given in the Appendix \ref{sec:proof_tene_consistency}.

\begin{figure}[t]
    \centering
    \hspace{-5.5em}
    \scalebox{0.6}{%
        \usetikzlibrary{arrows.meta, chains, positioning, calc}
\usetikzlibrary{shapes.geometric,shapes.arrows,fit,backgrounds,positioning}

\def\stride{2.1}



\begin{tikzpicture}
[
ff/.style    = { rectangle, draw=black, thick, 
    fill=ff_color, text width=10em, text centered,
    rounded corners, minimum height=4.5em },
embeddings/.style    = { rectangle, draw=black, thick, 
    fill=embeddings_color,
    rounded corners, minimum height=2em },
embed/.style    = { rectangle, draw=black, thick, 
    fill=emb_color, text width=4em, text centered,
    rounded corners, minimum height=2em },
invisible/.style = {rectangle},
four/.style    = { rectangle, draw=black, thick, 
    fill=fpca_color, text width=10em, text centered,
    rounded corners, minimum height=4.5em },
addnorm/.style    = { rectangle, draw=black, thick, 
    fill=add_norm_color, text width=10em, text centered,
    rounded corners, minimum height=2em },
den/.style    = { rectangle, draw=black, thick, 
    fill=dense_color, text width=10em, text centered,
    rounded corners, minimum height=2em },
outp/.style    = { rectangle, draw=black, thick, 
    fill=output_color, text width=10em, text centered,
    rounded corners, minimum height=2em },
line/.style     = { draw, thick, ->, shorten >=0pt, rounded corners=0.1cm},
strightline/.style     = { draw, thick, -, shorten >=0pt, rounded corners=0.1cm},
]
\definecolor{gray_bbox_color}{RGB}{243,243,244}
\definecolor{emb_color}{RGB}{252,224,225}
\definecolor{embeddings_color}{RGB}{232,204,205}
\definecolor{fpca_color}{RGB}{252,226,187}
\definecolor{add_norm_color}{RGB}{242,243,193}
\definecolor{ff_color}{RGB}{194,232,247}
\definecolor{output_color}{RGB}{203,231,207}
\definecolor{dense_color}{RGB}{220,223,240}


\matrix [column sep=1mm, row sep=5mm, xshift=-3cm] {
    & \node [text centered] (output) {$g_y({Y_l,Y^{l-1}})$};         & \\
    & \node [den] (dense) {Dense};                     & \\
    & \node [addnorm] (add2) at ([yshift=-6mm]dense.south) {Add \& Norm};       & \\
    & \node [ff] (ff) {Feed Forward};                  & \\
    & \coordinate (null2);                                & \\
    & \node [addnorm] (add) {Concat \& Norm};        & \\
    & \node [four] (fpca) {Fixed-Past-\\Causal-Attention};                & \\
    & \coordinate (null1);                                & \\
};

\node[below = 1.2cm of fpca] (embeddings) {Embeddings};
\node[below = 1mm of embeddings,invisible] (emb_inv) {+};
\node[left = 4mm of emb_inv,embed] (position) {Position};
\node[right = 4mm of emb_inv,embed] (value) {Value};
\path (position) -- (emb_inv) node [midway]{} -- (value) node [midway]{};

\begin{pgfonlayer}{background}
    \node[fit=(embeddings)(position)(value),embeddings](embed){};
    \coordinate (N1) at (embed.west|-fpca);
    \coordinate (N2) at (embed.east|-fpca);
    \node[inner xsep=0pt,inner ysep=22pt,fit=(add2)(fpca)(N1)(N2),fill=gray_bbox_color, line width=0.03cm,draw, rounded corners=0.2cm]{};
\end{pgfonlayer}

\node[below= 5mm of embed](input){$
    \begin{bmatrix}
        Y_{1},\cdots, Y_{l}
    \end{bmatrix}$
};

\begin{scope} [every path/.style=line]
    \path (dense)       --  (output);
    \path (add2)        --  (dense);
    \path (ff)          --  (add2);
    \path (add)         --  (ff);
    \path (null2)       --++  (-3,0) |- (add2);
    \path (fpca)     --  (add);
    \path (embed)  --  (fpca) coordinate[midway](null1) ;
    \path (null1)  --++  (-3,0) |- (add);
    \path (input) -- (embed);
\end{scope}

\node[below = 2mm of fpca,invisible] (split) {};
\begin{scope} [every path/.style=line]
    \path(split) --++ (1,0) -| ([xshift=-1.2cm]fpca.south);
    \path(split) --++ (-1,0) -| ([xshift=1.2cm]fpca.south);
\end{scope}

\node[above = 0.02mm of fpca.south](keys) {\footnotesize $\textbf{K}$};
\node[above right=0.02mm and 1cm of fpca.south](values) {\footnotesize $\textbf{V}$};
\node[above left=0.02mm and 1cm of fpca.south](queries) {\footnotesize $\smash{\textbf{Q}}$};


\matrix [column sep=1mm, row sep=5mm, xshift=3cm] {
    & \node [text centered] (2_output) {$g_y({\tilde{Y}_l, Y^{l-1}})$};         & \\
    & \node [den] (2_dense) {Dense};                     & \\
    & \node [addnorm] (2_add2) at ([yshift=-6mm] 2_dense.south) {Add \& Norm};       & \\
    & \node [ff] (2_ff) {Feed Forward};                  & \\
    & \coordinate (2_null2);                                & \\
    & \node [addnorm] (2_add) {Concat \& Norm};        & \\
    & \node [four] (2_fpca) {\textbf{Modified} Fixed-Past-\\Causal-Attention};                & \\
    & \coordinate (2_null1);                                & \\
};

\node[below = 1.2cm of 2_fpca] (2_embeddings) {Embeddings};
\node[below = 1mm of 2_embeddings,invisible] (2_emb_inv) {+};
\node[left = 3mm of 2_emb_inv,embed] (2_position) {Position};
\node[right = 3mm of 2_emb_inv,embed] (2_value) {Value};
\path (2_position) -- (2_emb_inv) node [midway]{} -- (2_value) node [midway]{};

\begin{pgfonlayer}{background}
    \node[fit=(2_embeddings)(2_position)(2_value),embeddings](2_embed){};
    \coordinate (2_N1) at (2_embed.west|-2_fpca);
    \coordinate (2_N2) at (2_embed.east|-2_fpca);
    \node[inner xsep=0pt,inner ysep=22pt,fit=(2_add2)(2_fpca)(2_N1)(2_N2),fill=gray_bbox_color, line width=0.03cm,draw, rounded corners=0.2cm]{};
\end{pgfonlayer}

\node[below= 5mm of 2_embed](2_input){$
    \begin{bmatrix}
        \smash{\tilde{Y}_{1},\cdots, \tilde{Y}_{l}}
    \end{bmatrix}$
};

\begin{scope} [every path/.style=line]
    \path (2_dense)       --  (2_output);
    \path (2_add2)        --  (2_dense);
    \path (2_ff)          --  (2_add2);
    \path (2_add)         --  (2_ff);
    \path (2_null2)       --++  (3,0) |- (2_add2);
    \path (2_fpca)     --  (2_add);
    \path (2_embed)  --  (2_fpca) coordinate[midway](2_null1) ;
    \path (2_null1)  --++  (3,0) |- (2_add);
    \path (2_input) -- (2_embed);
\end{scope}

\node[below = 2mm of 2_fpca,invisible] (2_split) {};
\begin{scope} [every path/.style=line]
    \path(2_split) --++ (1,0) -| ([xshift=-1.2cm]2_fpca.south);
    \path(2_split) --++ (-1,0) -| ([xshift=1.2cm]2_fpca.south);
\end{scope}

\node[above = 0.02mm of 2_fpca.south](2_keys) {\footnotesize $\tilde{\textbf{K}}$};
\node[above right=0.02mm and 1cm of 2_fpca.south](2_values) {\footnotesize $\tilde{\textbf{V}}$};
\node[above left=0.02mm and 1cm of 2_fpca.south](2_queries) {\footnotesize $\smash{\tilde{\textbf{Q}}}$};

\coordinate (midpoint) at ($(fpca)!0.5!(2_fpca)$);
\node[above] at (midpoint) {\footnotesize{$\textbf{K,V}$}};
\draw[line] (fpca) -- (2_fpca);

\end{tikzpicture}

%
    }
    \caption{The TREET architecture for $g_y$, with memory parameter $l$. It illustrated using a single sequence as an example. However, it is capable of parallel processing for sequences of length $L>l$, and in such cases, the number of outputs for the function will be $L-l+1$. Both transformer share the same weights. The FPCA and the modified FPCA are as elaborated in Section \ref{sec:fixed_past_causal_attention}.}
    \label{fig:treet_transformer_gy}
\end{figure}

\subsection{Algorithm and Implementation}\label{sec:implementation_and_algorithm}
\noindent
This section describes the TREET implementation. We present an overview scheme for estimation of TE and describe the algorithm. Then the TREET architecture is outlined. 
In practice, the TREET optimization boils down to gradient-based optimization of its parameters. Therefore, denote the TREET networks with $g_{\theta_y}, g_{\theta_{xy}}$ with corresponding parameters $\theta_{y}$ and $\theta_{xy}$ respectively.
For simplicity, we address mainly $g_{\theta_y}$ and $\DYT$, and afterwards explain how to extend all to $g_{\theta_{xy}}$ and $\DYXT$.

\subsubsection{Overview and Algorithm}
\noindent
The TREET algorithm follows an iterative joint optimization of $\hDYXT$ and $\hDYXT$ through iterative mini-batch gradient optimization.
The algorithm inputs are $l, \Dn$, which are the TE parameter and the dataset, respectively.
Every iteration begins with feeding mini-batch sized $m<n$ with sequences length $l$ in each model, followed by the calculation of both DV potentials \eqref{eq:dyt_and_dyxt}, that construct $\sTE(l)$ \eqref{eq:tene_objective_estimator}.
The calculated objective is then used for gradient-based optimization of the NN parameters.
The iterative process continues until a stopping criteria is met, typically defined as the convergence of $\tene(\Dn;l)$ within a specified tolerance parameter $\epsilon>0$. For evaluation, the sequence-wise TE is estimated over as many samples and then averaged to obtain the estimated TE.
The full pipeline for estimating $\hDYT(\Dn, g_{\theta_y})$ is presented in Fig. \ref{fig:tene_dy}, and the complete list of steps is given in Algorithm \ref{alg:tene}.

\begin{algorithm}[!t]
    \caption{TREET}
    \label{alg:tene}
    \textbf{Input:} Joint process samples $\Dn$; Observation length $l\in\NN$.\\
    \textbf{Output:} $\tene(\Dn;l)$ - TE estimation.
    \algrule
    \begin{algorithmic}[1]
    \State NNs initialization $g_{\theta_y}$, $g_{\theta_{xy}}$ with corresponding parameters $\theta_{y} ,\theta_{xy}$.
    \State \textbf{Step 1 -- Optimization:}
    \Repeat
    \State\hspace{-3mm}\parbox[t]{0.95 \linewidth}{%
        Draw a batch $B_m$: $m<n$ sub-sequences, length $L>l$ from $\Dn$, with reference samples $P_{\tilde{Y}}$ for each.
    }
    \State\hspace{-3mm}\parbox[t]{0.95 \linewidth}{%
        Compute both potentials $\hDYXT(\Bm, g_{\theta_{xy}})$, $\hDYT(\Bm, g_{\theta_{y}})$ via \eqref{eq:dyt_and_dyxt}.
    }
    \State\hspace{-3mm}Update parameters:\\
    \quad\quad$\theta_{xy} \leftarrow \theta_{xy} + \nabla_{\theta_{xy}}\hDYXT(\Bm, g_{\theta_{xy}})$\\
    \quad\quad$\theta_{y} \leftarrow \theta_{y} +\nabla_{\theta_{y}}\hDYT(\Bm, g_{\theta_{y}})$
    \Until{convergence criteria.}
    
    \State\textbf{Step 2 -- Evaluation:} Evaluate for a sub-sequence \eqref{eq:dyt_and_dyxt} and \eqref{eq:tene_estimator} to obtain $\tene(\Dn;l)$.
    \end{algorithmic}
\end{algorithm}

\begin{figure*}[t]
\begin{equation}\label{eq:qk_origin}
    \scalemath{0.9}{
    QK^\top = 
    \begin{bmatrix}
        {q}_{(t+l)}\cdot k_{(t)} & \ldots  & {q}_{(t+l)} \cdot {k}_{(t+l)} & -\infty & \ldots & -\infty \\
        -\infty & {q}_{(t+l+1)} \cdot k_{(t+1)} & \ldots & {q}_{(t+l+1)} \cdot {k}_{(t+l+1)} & -\infty & \vdots \\
        \vdots & -\infty & \ddots & & \ddots & -\infty \\
        -\infty & \ldots & -\infty & {q}_{(t+L)} \cdot k_{(t+L-l)} & \ldots & {q}_{(t+L)} \cdot {k}_{(t+L)} \\
    \end{bmatrix}}
\end{equation}
\end{figure*}
\begin{figure*}[t]
\begin{equation}\label{eq:qk_ref}
    \scalemath{0.9}{
    \widehat{QK^\top} = 
    \begin{bmatrix}
        \tilde{q}_{(t+l)} \cdot k_{(t)} & \ldots & \tilde{q}_{(t+l)} \cdot \tilde{k}_{(t+l)} & -\infty & \ldots & -\infty \\
        -\infty & \tilde{q}_{(t+l+1)} \cdot k_{(t+1)} & \ldots & \tilde{q}_{(t+l+1)} \cdot \tilde{k}_{(t+l+1)} & -\infty & \vdots \\
        \vdots & -\infty & \ddots & & \ddots & -\infty \\
        -\infty & \ldots & -\infty & \tilde{q}_{(t+L)} \cdot k_{(t+L-l)} & \ldots & \tilde{q}_{(t+L)} \cdot \tilde{k}_{(t+L)} \\
    \end{bmatrix}}.
\end{equation}
\end{figure*}

The DV representation, \eqref{eq:dv_represnetation}, suggest that the function $f$ is the same function for both terms that construct it, i.e. the weights are shared for both network propagation. Hence, our transformer in TREET, which constructed by \eqref{eq:dyt_and_dyxt}, is the same for both terms, for both DV representation. Exemplifying with $\hDYT$, both $g_y(Y^l)$ and $g_y(\tilde{Y}_l,Y^{l-1})$ use the same learning parameters, from positional encoding layers and attention to FF layers. The only difference between the two terms, is in how the attention mechanism operates, which essentially re-uses keys and values generated for the first term, to generate the second in \eqref{eq:dyt_expectation}. This model for $g_y$ is visualized in Fig. \ref{fig:treet_transformer_gy}.
Next, we elaborate about the proposed \textit{fixed past causal attention} that constructs the TREET and its variation, \textit{modified fixed past causal attention} which is required for the reference measurement distribution.

\subsubsection{Fixed Past Causal Attention }\label{sec:fixed_past_causal_attention}
To comply with \eqref{eq:dyt_and_dyxt} in Section \ref{subsec:estimator}, the model must avoid using future inputs relative to the prediction point and operate with a fixed sequence length of $l$. To enforce this constraint within the attention mechanism, a masking strategy introduced, termed as \textit{fixed past causal attention} (FPCA). Unlike the standard causal mask $M \in \RR^{L \times L}$ that simply prevents access to future values, FPCA employs a Toeplitz-like banded mask $M' \in \RR^{L \times L}$ defined as: 
\begin{equation}
    M'_{[i,j]} = 
    \begin{cases}
        1 & \text{if } j-i < l \text{ and } j \geq i \\
        -\infty & \text{otherwise}
    \end{cases}     
\end{equation}
This ensures that each attention query attends only to the current and previous $l - 1$ inputs, maintaining a fixed-length historical context. The FPCA is given by   \begin{equation}
    \mathsf{FPCA}:= \mathsf{softmax}( QK^\top \odot M') V
\end{equation}
FPCA matrix is visualized in \eqref{eq:qk_origin}.
Note that only the last $L-l+1$ results from the transformer outputs sequence are used, in order to keep the past information fix to length $l$. Hence, the complexity of the proposed attention operation is $\cO(L ld_o)$.

\subsubsection{Reference Sampling with FPCA}\label{sec:reference_sampling_attention}
\noindent
This section describes the calculation of $g_y(\tilde{Y}_l,Y^{l-1})$ \eqref{eq:dyt_and_dyxt_a}.
Denote the scoring value after the softmax operation of FPCA for query $q_i$ with key $k_j$ as 
\begin{equation}
    \cR_{(i,j)}:=\frac{e^{q_i \cdot k_j}}{\sum_{m=i-l}^{i}e^{q_i \cdot k_m}},\quad j\leq i.
\end{equation}
Since FPCA with memory parameter $l$ is used, the attention output at time $t$ can be written as $\FPCA_t=\cR_{(t,t-l)}v_{t-l}+\cR_{(t,t-l+1)}v_{t-l+1}+\ldots+\cR_{(t,t)}v_{t}$.
In order to calculate $g_y(\tilde{Y}_t,Y^{t-1}_{t-l})$, FPCA should use information from the keys and values that were used to calculate $g_y(Y^t_{t-l})$. To that end, we modify the FPCA mechanism as follows:
\begin{equation}
    \tilde{\cR}_{(i,j)}:=
    \begin{cases}
        \frac{e^{ \tilde{q}_i \cdot k_j }}{e^{ \tilde{q}_i \cdot \tilde{k}_i } + \sum_{m=i-l}^{i-1} e^{ \tilde{q}_i \cdot k_m} }
        &, \quad j < i \\
        \frac{e^{ \tilde{q}_i \cdot\tilde{k}_i }}{e^{ \tilde{q}_i \cdot \tilde{k}_i } + \sum_{m=i-l}^{i-1}e^{\tilde{q}_i \cdot k_m}} & ,\quad  i=j.
    \end{cases}
\end{equation}

In this case, the modified FPCA for time $t$ can be written as $\mathsf{Modified}\text{-}\FPCA_t=\tilde{\cR}_{(t,t-l)}v_{t-l}+\tilde{\cR}_{(t,t-l+1)}v_{t-l+1}+\ldots+\tilde{\cR}_{(t,t)}\tilde{v}_{t}$.
Summarizing, the second term \eqref{eq:dyt_and_dyxt_a} generated by the modified FPCA, contains all keys and values of the relative past and query, key and value for the current present, which are a function of the reference distribution.
The dot-product matrix between queries and keys is visualized in \eqref{eq:qk_ref},
for $l \leq L$, and modified FPCA is written as $\mathsf{Modified}\text{-}\mathsf{FPCA}= \mathsf{softmax}( \hat{QK^\top} \odot M') V$. 
FPCA and modified FPCA immediately extend to multi-head setting.

In our implementation, the reference input $\tilde{Y}$, for the second term in \eqref{eq:dyt_and_dyxt_a} are drawn from the uniform measure on the bounding box of the current batch of $Y$ samples, while theoretically, our method allows to draw from any positive continuous distribution measure; for further details check Appendix \ref{sec:proof_tene_consistency}.
The implementation of $\hDYXT(\Dn, g_{xy})$ \eqref{eq:dyt_and_dyxt_b} is obtained by concatenating the $X^l$ values with the corresponding $Y^l$ or $(\tilde{Y}_l, Y^{l-1})$ values for both the FPCA and the modified FPCA, respectively.


\section{Optimization of Estimated Transfer Entropy}\label{sec:optimization_te}
\noindent
Many data-driven problems reduce to steering an input distribution to maximize information flow, analogous to reinforcement learning, where a policy is learned to maximize expected reward. Thus, applications can leverage TE optimization by controlling the distributions of processes $\XX$ and $\YY$, either independently or jointly. Communication channel capacity offers a canonical example: it equals the DI rate and, by Section \ref{sec:te_and_di_relations}, can be estimated via TE. Because capacity embodies the fundamental limit of information flow, demonstrating TREET optimization with on this task provides a proof-of-concept readily generalizable to other TE-based control problems.

\begin{figure}[t]
    \centering
    \hspace{-2em}
    \scalebox{0.6}{%
        \usetikzlibrary{arrows.meta, chains, positioning, calc}
\def\stride{1.7}
\tikzstyle{block_big} = [rectangle, minimum width=\stride*4 cm, minimum height=1.5cm, align=center, draw=black, fill=black!5]
\tikzstyle{block_small} = [rectangle, minimum width=\stride cm, minimum height=1.2cm, align=center, draw=black, fill=white]

\tikzstyle{invisible} = [rectangle]


\begin{tikzpicture}[
line/.style     = { draw, thick, ->, shorten >=0pt },
]
\definecolor{ndg_color}{RGB}{232,247,232}

\def\x{0}
\node[block_small, , anchor=north west] (xt1) at (\x*\stride, -2.5) {
    $\begin{bmatrix}
        X_{t-L} \\
        0
    \end{bmatrix}$
};

\def\x{1}
\node[block_small, , anchor=north west] (xt2) at (\x*\stride, -2.5) {
    $\begin{matrix}
        \cdots
    \end{matrix}$
};
\node[above=0.3cm] at (xt2.north) {$\cdots$}; 

\def\x{2}
\node[block_small, , anchor=north west] (xt3) at (\x*\stride, -2.5) {
    $\begin{bmatrix}
        X_{t-1} \\
        0
    \end{bmatrix}$
};

\def\x{3}
\node[block_small, , anchor=north west] (xt4) at (\x*\stride, -2.5) {
    $\begin{bmatrix}
        0 \\
        U_t
    \end{bmatrix}$
};

\node[block_small, , anchor=north west] (xt) at (\x*\stride, 2) {
    $X_t$
};

\node[block_big, anchor=north west, rounded corners=0.3cm, fill=ndg_color] (model) at (0,0) {NDG\\Transformer};

\draw [line] (xt1.north) -- (xt1|-model.south);
\draw [line] (xt3.north) -- (xt3|-model.south);
\draw [line] (xt4.north) -- (xt4|-model.south);

\draw [line] (xt|-model.north) -- (xt);

\draw [-] (xt.north) 
    to[out=90,in=95, looseness=1.5] ($(xt.east)+(1,0.5)$) 
    to[out=0,in=0, looseness=0] ($(model.east)+(1,.4)$);
\draw [-] ($(model.east)+(1,.4)$) -- ($(model.east) + (0.3,-.4)$);
\draw [line, bend left=45] ($(model.east) + (1,-.4)$)
    to[out=0,in=0, looseness=-1] ($(xt4.east)+(1,-0.5)$) 
    to[out=90,in=90] (xt3.south);

\end{tikzpicture}

%
    }
    \caption{The recursive process for the NDG with transformers. If feedback presents, the input includes the past channel output, concatenated with the corresponding value $X_i$ at time $i$.}
    \label{fig:ndg_transformers}
\end{figure}

\begin{remark}[Channel capacity]\label{remark:channel_capacity}
Consider channels with and without feedback links from the channel output back to the encoder. The feedforward capacity of a channel sequence $\left\{ P_{Y^n \| X^n}\right \}$, for $n\in\NN$ is
\begin{equation}
    C_\ff = \lim_{n\to\infty}\sup_{P_{X^n}}\frac{1}{n}\sI(X^n;Y^n),
\end{equation}
and the feedback capacity is
\begin{equation}
    C_\fb = \lim_{n\to\infty}\sup_{P_{X^n\|Y^{n-1}}}\frac{1}{n}\sI(X^n\to Y^n).
\end{equation}
The achievability of the capacities is further discussed in \cite{dobrushin1963general}, \cite{tatikonda2008capacity}. \cite{massey1990causality} showed that for non-feedback scenario, the optimization problem over $P_{X^n\|Y^n}$ can be translated to $P_{X^n}$, which support the use of DI rate for both optimization problems \cite{tsur2023neural}. Under some conditions TE can estimate the DI rate, as presented in Section \ref{sec:te_and_di_relations}.
\end{remark}

Focusing on channel capacity, 
and assuming that the input distribution sampling mechanism is controllable, we propose an algorithm for the optimization of estimated TE with respect to the input generator.
We refer to this model as the \textit{neural distribution generator} (NDG).
The estimated TE optimization methodology is inspired by the proposed methods from \cite{tsur2023neural}. However, the adaptation to transformer architectures considers a different implementation of the proposed scheme.
As the TREET estimates TE from samples, the NDG is defined as a generative model of the input distribution samples, and is optimized with the goal of maximizing the downstream estimated TE.
\begin{lemma}[Optimal TE]\label{lemma:optimal_te}
    Let $(\XX, \YY)$ be jointly stationary processes, and the TE with memory parameter $l\in\NN$. Then, the maximal TE, $\sTE^\star(l)$, by $\XX$ is
    \begin{equation}
        \sTE^\star(l):= \sup_{P_{X^l}} \sI(X^l;Y_l|Y^{l-1}).
    \end{equation}
\end{lemma}
The proof of the lemma is given in Appendix \ref{sec:proof_optimal_te}.
The lemma suggests that NDG with input sequence length $l$ is enough to achieve maximum TE with memory parameter $l$, for independently controlling the distribution of $\XX$.

\begin{figure}[t]
    \centering
    \hspace{-2.5em}
    \scalebox{0.55}{%
        \usetikzlibrary{arrows.meta, chains, positioning, calc}

\tikzstyle{block_big} = [rectangle, rounded corners, minimum width=3cm, minimum height=3cm, align=center, draw=black, fill=black!5]
\tikzstyle{block_small} = [rectangle, rounded corners, minimum width=1.5cm, minimum height=1.5cm, align=center, draw=black, fill=black!5]

\tikzstyle{block_D} = [rectangle, rounded corners, minimum width=2.5cm, minimum height=1.4cm, align=center, draw=black, fill=black!5]
\tikzstyle{block_E} = [rectangle, rounded corners, minimum width=0.8cm, minimum height=0.8cm, align=center, draw=black, fill=black!5]
\tikzstyle{block_C} = [rectangle, minimum width=1cm, minimum height=0.01cm, draw=white, fill=white]

\tikzstyle{invisible} = [rectangle]


\begin{tikzpicture}
\definecolor{transformer_color}{RGB}{232,232,247}
\definecolor{ndg_color}{RGB}{232,247,232}


\node (input_ui) at (0,0) {};
\node [block_big, fill=ndg_color] (ndg) at (3, 0) {\large NDG \\ \large Transformer \\ \large $\phi$};
\node [block_big] (system) at (7.5, 0) {\large Channel\\ $P_{Y_i|X^{i}_{i-l},Y^{i-1}_{i-l}}$};
\node [block_big, fill=transformer_color] (transformer) at (12, 0) {\large TREET \\ \large Transformer \\ \large $g_{\theta_{y}},g_{\theta_{xy}}$};

\node (ndg_additional) at (1, 2) {$X_{i-l}^{i-1}$};
\node (system_addition) at (9.8, 2) {$X_{i-l}^{i-1}, Y_{i-l}^{i-1}$};

\node [block_big] (loss) at (7.5, -4) {\large DV \\ \large Loss};

\draw [arrows = {-Latex[width=7pt, length=7pt]}] (input_ui) -- node[above=1mm] {$U_i$} (ndg.west);

\draw [arrows = {-Latex[width=7pt, length=7pt]}] (ndg) -- node[above=1mm] {$X_i$} (system.west);

\draw [arrows = {-Latex[width=7pt, length=7pt]}] (system) -- node[above=1mm] {$Y_i,X_i$} (transformer.west);

\draw [arrows = {-Latex[width=7pt, length=7pt]}] ([xshift=-15] transformer.south) -- ([xshift=-15, yshift=-13] transformer.south) |- ([yshift=15] loss.north) --  (loss.north);

\draw [dashed] (loss.east) -- node[above=1mm] {$\nabla g_{\theta_y}, \nabla g_{\theta_{xy}}$} ([xshift=70]loss.east);
\draw [dashed] ([xshift=70]loss.east) -- ++ (.5,-.5);
\draw [arrows = {-Latex[width=7pt, length=7pt]}, dashed] ([xshift=170]loss.west) -| ([xshift=15] transformer.south);

\draw [dashed] (loss.west) -- node[above=1mm] {$ \nabla \phi$} ([xshift=-35]loss.west);
\draw [dashed] ([xshift=-35]loss.west) -- ++(-.5,-.5);
\draw [arrows = {-Latex[width=7pt, length=7pt]}, dashed] ([xshift=-50] loss.west) -|   (ndg.south);

\draw [arrows = {-Latex[width=7pt, length=7pt]}] ([xshift=-5] ndg_additional.south) |- node[above right=.25mm] { $\star$} ([yshift=20] ndg.west);

\draw [arrows = {-Latex[width=7pt, length=7pt]}] ([xshift=3] system_addition.south) |- ([yshift=20] transformer.west);

\end{tikzpicture}
%
    }
    \caption{Complete system for estimating and optimizing TREET with NDG while Altering between the models to train on. ($\star$)  If feedback presents, past channel output realizations included.}
    \label{fig:tene_with_ndg_opt}
\end{figure}

The NDG calculates a sequence of channel input $X^{l}$ through the mapping 
\begin{equation}
    \NDG:(U_i,Z^{i-1}_{i-l})\rightarrow X_{i}^{\phi}, \qquad i=1,\ldots,n, 
\end{equation}
where $U_i$ is the random noise drawn from $P_U\in\lebmeas(\cU), \cU\subset\RR^{d_x}$, which cause the stochasticity, $Z_{i-l}^{i-1}$ are the past observation of the generated process created by the model, and the channel corresponding outputs if feedback exists. $\NDG$ is the parametric NDG mapping with parameters $\phi\in\Phi$.
By the functional representation lemma \cite{el2011network} and the restated lemma in \cite{tsur2023neural}, the distribution of $X$ is achievable from an NN function.
After $l$ iterations, with $Z_{i-l}^{i-1}$ storing the information of previous iterations, the NDG generates the whole sequence.

Transformers need access to the whole sequence at once, in contrast to RNNs where a single state can theoretically represent the past sequence.
Thus, the input sequence to the NDG with transformer is created via past outputs of the transformer itself (and corresponding channel outputs if feedback exists) as depicted in Fig. \ref{fig:ndg_transformers}, and the past observations are taken without gradients to prevent backpropagation through iterations.
In our implementation, the input convention for each time step is a concatenated vector of $[X_i, U_i ]^\top$ to maintain the input structure of samples and random noise for every projection in the network. In relative history time steps, the noise is replaced with a zero vector, and for the present time, the input $X_i$ is replaced with a zero vector.

\begin{algorithm}[t!]
\caption{Continuous TREET optimization}
\label{alg:cap_est_ndt}
\textbf{Input:} Continuous sequence-to-sequence system $\mathcal{S}$; Observation length $l\in\NN$.\\
\textbf{Output:} $\tene^\star(U^n; l)$, optimized NDG.
\algrule
\begin{algorithmic}[1]
\State 
NNs initialization $g_{\theta_y}$, $g_{\theta_{xy}}$ and $\NDG$ with corresponding parameters $\theta_{y} ,\theta_{xy}, \phi$.
\Repeat
    \State Draw noise $U^m$, $m<n$. 
    \State Compute batch $\Bphi$ sized $m$ using NDG, $\mathcal{S}$
    \If{training TREET}
    \State \parbox[t]{0.8 \linewidth}{%
        Perform TREET optimization - Step 1 in Algorithm \ref{alg:tene}.
    }
    \Else \hspace{0.15cm}Train NDG
    \State \parbox[t]{0.85 \linewidth}{%
        Compute $\tene(\Bphi, g_{\theta_{y}},g_{\theta_{xy}},\NDG; l)$ using \eqref{eq:tene_estimator}. 
    }
    \State Update NDG parameters: \\
    \hspace{\algorithmicindent}$\quad \phi \leftarrow \phi + \nabla_{\phi}\tene(\Bphi, g_{\theta_{y}} ,g_{\theta_{xy}},\NDG; l)$
    \EndIf
\Until{convergence criteria.}
\State Draw $U^m$ to produce $l$ length sequence and evaluate $\tene(\Dn^\phi; l)$. \\
\Return $\tene^\star(U^n; l)$, optimized NDG.
\end{algorithmic}
\end{algorithm}

To estimate and optimize the TE at once, we jointly training the TREET and the NDG. As described in Algorithm \ref{alg:cap_est_ndt} and Fig. \ref{fig:tene_with_ndg_opt}, in each iteration only one of those model is updated by maximizing each DV potentials, $\hDYT, \hDYXT$, for TREET model and by maximizing $\tene$ for NDG model. 
The entire pipeline of one iteration creates a sequence length $l$ from the NDG, by iterating it $l$ times with some initiated zero values - which creates the dataset, $\Dn^\phi=(X^{\phi,n},Y^{\phi,n})$. Afterwards, feeding each sample sequence from the dataset to TREET for achieving the corresponding networks' outputs as mentioned back in Algorithm \ref{alg:tene}, which constructs the loss that generate gradients backward according to each model.
Next, we demonstrate the power of the proposed method for estimating the channel capacity.


\section{Experimental Results}\label{sec:results}
\noindent
Evaluation of TREET comprises four interconnected experiments that build upon one another to tell a unified narrative: first, assessing core TE estimation accuracy; next, harnessing TREET for optimization of information flow; then demonstrating TREET’s built-in distribution-estimation utility; and finally applying it as a feature-analysis tool on real-world clinical data, illustrating how foundational performance extends seamlessly to practical applications.

Section~\ref{sec:exp_benchmark} benchmarks TE estimation on synthetic long-memory data, extending \cite{zhang2019itene} and comparing TREET against TENE and Copnet.  
Section~\ref{sec:exp_capacity_opt} integrates TREET with NDG to optimize communication channel capacity, compared with DINE, the RNN-based DI estimator \cite{tsur2023neural}, and Section~\ref{sec:exp_capacity_memory} further analyses the optimization and estimation memory performance. While estimation experiments are about continuous spaces of distributions, discrete spaces can easily be applied to the algorithm of TREET. 
Section~\ref{sec:exp_density} applies TREET to PDF estimation of memory processes, demonstrating adaptability to continuous-space distribution tasks. 
Section~\ref{sec:exp_real_world} presents a real-world case study using TREET for feature analysis on an Apnea patients dataset. 

All experiments consider a TREET network with a single FPCA layer follows by a single FF layer, and for the optimization procedure the NDG consists of transformer with a single causal-attention layer and a single FF layer. Further details about the implementation\footnote{The code implementation can be found at our Github repository \href{https://github.com/omerlux/TREET}{\underline{https://github.com/omerlux/TREET}}} are provided in Appendix \ref{sec:appendix_implementation}.

\addtolength{\tabcolsep}{-6pt}    
\begin{table}[t]
    \centering
    \newcommand{\benchmarkrowlen}{350pt}
    \resizebox{0.48\textwidth}{!}{%
\newcommand{\GT}[1]{%
    \ifnum#1=-3 0.829\fi%
    \ifnum#1=-2 0.811\fi%
    \ifnum#1=-1 0.699\fi%
    \ifnum#1=0 0.415\fi%
    \ifnum#1=1 0.132\fi%
    \ifnum#1=2 0.019\fi%
    \ifnum#1=3 0.001\fi%
}

    \begin{tabular}{cc|ccccccc}
        \toprule
        $\lambda$ &  & \textbf{-3} & \textbf{-2} & \textbf{-1} & \textbf{0} & \textbf{1} & \textbf{2} & \textbf{3} \\ 
        \midrule

        \multicolumn{2}{c|}{\diaghead{\theadfont Ground Ground Truth}%
        {Model, $l$}{Ground$\quad$\\Truth$\quad$}}
        & \GT{-3} & \GT{-2} & \GT{-1} & \GT{0} & \GT{1} & \GT{2} & \GT{3} \\ 
        \midrule
        \multirow{8}{*}{$\quad$\textbf{\rotatebox[origin=c]{90}{TREET}}} 
        & $1$  & \diffColor{0.812}{\GT{-3}} & \diffColor{0.792}{\GT{-2}} & \diffColor{0.667}{\GT{-1}} & \diffColor{0.395}{\GT{0}} & \diffColor{0.126}{\GT{1}} & \diffColor{0.016}{\GT{2}} & \diffColor{-0.000}{\GT{3}} \\ 
        & $2$  & \diffColor{0.826}{\GT{-3}} & \diffColor{0.796}{\GT{-2}} & \diffColor{0.681}{\GT{-1}} & \diffColor{0.392}{\GT{0}} & \diffColor{0.117}{\GT{1}} & \diffColor{0.014}{\GT{2}} & \diffColor{0.000}{\GT{3}} \\ 
        & $4$  & \diffColor{0.825}{\GT{-3}} & \diffColor{0.798}{\GT{-2}} & \diffColor{0.682}{\GT{-1}} & \diffColor{0.393}{\GT{0}} & \diffColor{0.115}{\GT{1}} & \diffColor{0.013}{\GT{2}} & \diffColor{0.000}{\GT{3}} \\ 
        & $7$  & \diffColor{0.820}{\GT{-3}} & \diffColor{0.799}{\GT{-2}} & \diffColor{0.679}{\GT{-1}} & \diffColor{0.405}{\GT{0}} & \diffColor{0.121}{\GT{1}} & \diffColor{0.015}{\GT{2}} & \diffColor{0.001}{\GT{3}} \\  
        & $9$  & \diffColor{0.815}{\GT{-3}} & \diffColor{0.802}{\GT{-2}} & \diffColor{0.686}{\GT{-1}} & \diffColor{0.403}{\GT{0}} & \diffColor{0.126}{\GT{1}} & \diffColor{0.017}{\GT{2}} & \diffColor{0.001}{\GT{3}} \\ 
        & $19$ & \diffColor{0.824}{\GT{-3}} & \diffColor{0.805}{\GT{-2}} & \diffColor{0.690}{\GT{-1}} & \diffColor{0.405}{\GT{0}} & \diffColor{0.128}{\GT{1}} & \diffColor{0.017}{\GT{2}} & \diffColor{0.001}{\GT{3}} \\ 
        & $49$ & \diffColor{0.829}{\GT{-3}} & \diffColor{0.811}{\GT{-2}} & \diffColor{0.694}{\GT{-1}} & \diffColor{0.409}{\GT{0}} & \diffColor{0.128}{\GT{1}} & \diffColor{0.018}{\GT{2}} & \diffColor{0.001}{\GT{3}} \\ 
        & $99$ & \diffColor{0.829}{\GT{-3}} & \diffColor{0.810}{\GT{-2}} & \diffColor{0.693}{\GT{-1}} & \diffColor{0.410}{\GT{0}} & \diffColor{0.115}{\GT{1}} & \diffColor{0.017}{\GT{2}} & \diffColor{0.001}{\GT{3}} \\ 
        \midrule
        \multirow{8}{*}{$\quad$\textbf{\rotatebox[origin=c]{90}{TENE}}} 
        & $1$  & \diffColor{0.823}{\GT{-3}} & \diffColor{0.807}{\GT{-2}} & \diffColor{0.696}{\GT{-1}} & \diffColor{0.416}{\GT{0}} & \diffColor{0.126}{\GT{1}} & \diffColor{0.014}{\GT{2}} & \diffColor{0.000}{\GT{3}} \\ 
        & $2$  & \diffColor{0.814}{\GT{-3}} & \diffColor{0.782}{\GT{-2}} & \diffColor{0.688}{\GT{-1}} & \diffColor{0.390}{\GT{0}} & \diffColor{0.115}{\GT{1}} & \diffColor{0.013}{\GT{2}} & \diffColor{0.000}{\GT{3}} \\ 
        & $4$  & \diffColor{0.430}{\GT{-3}} & \diffColor{0.760}{\GT{-2}} & \diffColor{0.602}{\GT{-1}} & \diffColor{0.382}{\GT{0}} & \diffColor{0.119}{\GT{1}} & \diffColor{0.013}{\GT{2}} & \diffColor{-0.001}{\GT{3}} \\ 
        & $7$  & \diffColor{999}{\GT{-3}} & \diffColor{999}{\GT{-2}} & \diffColor{0.698}{\GT{-1}} & \diffColor{0.354}{\GT{0}} & \diffColor{0.090}{\GT{1}} & \diffColor{0.013}{\GT{2}} & \diffColor{-0.000}{\GT{3}} \\  
        & $9$  & \diffColor{999}{\GT{-3}} & \diffColor{999}{\GT{-2}} & \diffColor{524.551}{\GT{-1}} & \diffColor{0.359}{\GT{0}} & \diffColor{0.091}{\GT{1}} & \diffColor{0.014}{\GT{2}} & \diffColor{-0.000}{\GT{3}} \\ 
        & $19$ & \diffColor{999}{\GT{-3}} & \diffColor{999}{\GT{-2}} & \diffColor{999}{\GT{-1}} & \diffColor{1.796}{\GT{0}} & \diffColor{0.038}{\GT{1}} & \diffColor{0.021}{\GT{2}} & \diffColor{0.000}{\GT{3}} \\ 
        & $49$ & \diffColor{-999}{\GT{-3}} & \diffColor{-999}{\GT{-2}} & \diffColor{-999}{\GT{-1}} & \diffColor{999}{\GT{0}} & \diffColor{0.027}{\GT{1}} & \diffColor{0.010}{\GT{2}} & \diffColor{-0.002}{\GT{3}} \\ 
        & $99$ & \diffColor{999}{\GT{-3}} & \diffColor{999}{\GT{-2}} & \diffColor{999}{\GT{-1}} & \diffColor{999}{\GT{0}} & \diffColor{0.102}{\GT{1}} & \diffColor{-0.002}{\GT{2}} & \diffColor{-0.002}{\GT{3}} \\ 
        \midrule
        \multirow{8}{*}{$\quad$\textbf{\rotatebox[origin=c]{90}{Copnet}}}  
        & $1$  & \diffColor{0.835}{\GT{-3}} & \diffColor{0.810}{\GT{-2}} & \diffColor{0.688}{\GT{-1}} & \diffColor{0.397}{\GT{0}} & \diffColor{0.111}{\GT{1}} & \diffColor{-0.000}{\GT{2}} & \diffColor{-0.022}{\GT{3}} \\ 
        & $2$  & \diffColor{0.819}{\GT{-3}} & \diffColor{0.786}{\GT{-2}} & \diffColor{0.676}{\GT{-1}} & \diffColor{0.377}{\GT{0}} & \diffColor{0.106}{\GT{1}} & \diffColor{-0.004}{\GT{2}} & \diffColor{-0.017}{\GT{3}} \\ 
        & $4$  & \diffColor{0.900}{\GT{-3}} & \diffColor{0.864}{\GT{-2}} & \diffColor{0.747}{\GT{-1}} & \diffColor{0.469}{\GT{0}} & \diffColor{0.193}{\GT{1}} & \diffColor{0.087}{\GT{2}} & \diffColor{0.079}{\GT{3}} \\ 
        & $7$  & \diffColor{1.297}{\GT{-3}} & \diffColor{1.268}{\GT{-2}} & \diffColor{1.135}{\GT{-1}} & \diffColor{0.903}{\GT{0}} & \diffColor{0.649}{\GT{1}} & \diffColor{0.571}{\GT{2}} & \diffColor{0.575}{\GT{3}} \\  
        & $9$  & \diffColor{1.703}{\GT{-3}} & \diffColor{1.679}{\GT{-2}} & \diffColor{1.558}{\GT{-1}} & \diffColor{1.357}{\GT{0}} & \diffColor{1.106}{\GT{1}} & \diffColor{1.054}{\GT{2}} & \diffColor{1.053}{\GT{3}} \\ 
        & $19$ & \diffColor{4.862}{\GT{-3}} & \diffColor{4.834}{\GT{-2}} & \diffColor{4.750}{\GT{-1}} & \diffColor{4.574}{\GT{0}} & \diffColor{4.431}{\GT{1}} & \diffColor{4.428}{\GT{2}} & \diffColor{4.432}{\GT{3}} \\ 
        & $49$ & \diffColor{17.958}{\GT{-3}} & \diffColor{17.936}{\GT{-2}} & \diffColor{17.859}{\GT{-1}} & \diffColor{17.779}{\GT{0}} & \diffColor{17.622}{\GT{1}} & \diffColor{17.669}{\GT{2}} & \diffColor{17.687}{\GT{3}} \\ 
        & $99$ & \diffColor{43.625}{\GT{-3}} & \diffColor{43.572}{\GT{-2}} & \diffColor{43.547}{\GT{-1}} & \diffColor{43.421}{\GT{0}} & \diffColor{43.356}{\GT{1}} & \diffColor{43.445}{\GT{2}} & \diffColor{43.442}{\GT{3}} \\ 

        \bottomrule
    \end{tabular}%


        }
    \caption{
    TE estimation extended benchmark 
    Comparing \treetext{}, TENE and Copnet. 
    The process is given by \eqref{eq:tenedata_flipping_process} and the true TE can be calculated for varying $\lambda$ \cite{zhang2019itene}. 
    The color intensity signifies the extent of deviation from the ground truth.
    The value is constant for any order - $l \geq 1$ since $\sTE(l) = \sTE(1)$.
    Our proposed benchmark is an extended version of the one presented in \cite{zhang2019itene} to include variable length input to calculate different orders of TE.
    In the given benchmark, $\rho$ is set to 0.9, while the $\lambda$ and $l$ parameters are varied.
    }
    \label{tab:te_benchmark}
\end{table}
\addtolength{\tabcolsep}{6pt}

\subsection{Transfer Entropy Benchmark}\label{sec:exp_benchmark}
\noindent
In order to present TREET as robust estimator, we extend the benchmark from \cite{zhang2019itene} to handle higher orders of TE.
Considering the following system:
\begin{equation}\label{eq:tenedata_flipping_process}
    Y_t =
    \begin{cases}
        Z_{t}, & \text{if } Y_{t-1} < \lambda, \\
        \rho X_{t-1} + \sqrt{1-\rho^2}Z_{t}, & \text{else,}
    \end{cases}
\end{equation}
where $\lambda \in \mathbb{R}$ is the process threshold, $\rho \in [0,1]$ determines the dependency between $X_{t-1}$ and $Z_{t}$, for ${X_t, Z_t}$ i.i.d. $\mathcal{N}(0, 1)$. The following equality holds - $\sTE(l)=\sTE(1), \quad \forall l\geq 1$ \cite{zhang2019itene}.
Comparing TREET with Copnet and TENE.

Copnet \cite{ma2019estimating} is empirical-density based non-parametric method that estimates the TE by four different values of copula entropy, which each one equals to a different MI value. We edited their version of estimator to handle conditioning on longer context length than one. Another estimator is TENE \cite{zhang2019itene}, parametric estimator, which utilizes the DV representation for estimation of two MI terms which are non-conditional, thus is prone to break with higher orders of TE since the input dimension to the DV optimization function will increase. For a fair comparison, 
the experiment was recreated using NN architectures with similarly sized parameters for both TENE and TREET and a training batch size of $1024$.

As shown in Table \ref{tab:te_benchmark}, TREET achieves performance comparable to TENE and Copnet for TE parameter $l = 1$, and process parameter $\rho=0.9$. However, as $l$ increases, both TENE and Copnet struggle to estimate TE accurately due to the increasing input dimensionality associated with longer context windows. In contrast, TREET effectively handles the additional temporal information, even for $l = 99$. This highlights the advantages of TREET's architecture in accommodating time dependencies.

\raggedbottom

\subsection{Capacity Estimation for Finite Memory Processes}\label{sec:exp_capacity_opt}
\begin{figure*}[t]
  \begin{subfigure}[t]{0.32\textwidth}
    \centering
    \scalebox{0.72}{%
        \input{Figures/awgn_dim}%
    }
    \caption{Capacity of AWGN (variate dimension).}
    \label{fig:ndt_awgn_dim}
  \end{subfigure}
  \hfill
  \begin{subfigure}[t]{0.32\textwidth}
    \centering
    \scalebox{0.72}{%
        \input{Figures/awgn_db}%
    }
    \caption{Capacity of AWGN (variate SNR).}
    \label{fig:ndt_awgn_db}
  \end{subfigure}
  \hfill
  \begin{subfigure}[t]{0.32\textwidth}
    \centering
    \scalebox{0.72}{%
        \input{Figures/gma1_fb_db}%
    }
    \caption{Feedback Capacity of Gaussian MA(1).}
    \label{fig:ndt_gma1_db_fb}
  \end{subfigure}
  \begin{subfigure}[t]{0.32\textwidth}
    \centering
    \scalebox{0.72}{%
        \input{Figures/gma1_ff_db}%
    }
    \caption{Feedforward Capacity of Gaussian MA(1).}
    \label{fig:ndt_gma1_db_ff}
  \end{subfigure}
  \hfill
  \begin{subfigure}[t]{0.32\textwidth}
    \centering
    \scalebox{0.72}{%
        \input{Figures/gar1_fb_db}%
    }
    \caption{Feedback Capacity of Gaussian AR(1).}
    \label{fig:ndt_gar1_db_fb}
  \end{subfigure}
  \hfill
  \begin{subfigure}[t]{0.32\textwidth}
    \centering
    \scalebox{0.72}{%
        \input{Figures/gar1_ff_db}%
    }
    \caption{Feedforward Capacity of Gaussian AR(1).}
    \label{fig:ndt_gar1_db_ff}
  \end{subfigure}

  \caption{Channel capacity estimations. 
  (a) Capacity of AWGN channel as a function of vector dimension for the 0 dB case. 
  (b) Capacity of AWGN channel as a function of SNR. 
  (c) Feedback capacity of Gaussian MA(1) channel with variate SNR. 
  (d) Feedforward capacity of Gaussian MA(1) channel with variate SNR.
  (e) Feedback capacity of Gaussian AR(1) channel with variate SNR.
  (f) Feedforward capacity of Gaussian AR(1) channel with variate SNR.
  $P$ represents the power constraint and the noise parameter is $\sigma^2$.}
  \label{fig:combined_capacity}
\end{figure*}
\raggedbottom
As shown in Section \ref{sec:te_and_di_relations}, under some conditions TE is equal to the DI rate and converges to it. DINE \cite{tsur2023neural} proved that it can estimate a capacity of stationary channels by optimizing the DI rate estimator and the input distribution of the channel $P_X$, as Remark \ref{remark:channel_capacity} mentions about channel capacities. Our experiments have shown that for memory-less channels and for memory channels with and without feedback, TREET (for an appropriate $l$ order) can approximate the DI rate, which is estimates the capacity, with the joint optimization procedure of the input distribution (NDG) and TE estimation. All of the results are presented in Fig. \ref{fig:combined_capacity}.
Important to note that channel input constraints are essential for ensuring that the transmitted signals are well-suited to the channel's characteristics and limitations. 
In this experiment, the power constraint on the input signal of the channel, which is implemented via normalizing a batch of samples to a certain statistics according to the power constraint. 
While we present results on continuous-valued processes, TREET can be optimal for discrete data, using methods, such as reinforcement learning optimization \cite{tsur2023data}.
\raggedbottom

\begin{figure}[t]
    \centering
    \scalebox{0.72}{%
        \input{Figures/ndt.tex}%
    }
    \caption{NDG input noise and output $X$, $0$ SNR case.}
    \label{fig:ndt_input_output}
\end{figure}

\begin{figure*}[t]
  \begin{subfigure}[t]{1\textwidth}
    \centering
    \scalebox{0.72}{%
\pgfplotsset{colormap/blackwhite}
\begin{tikzpicture}[trim axis left, trim axis right]

\definecolor{darkgray176}{RGB}{176,176,176}

\begin{axis}[
colorbar,
colorbar style={
    ylabel={Attention weight},
    ylabel style={
        yshift=-.9cm,  
    },
    xshift=.4cm,
    width=.2cm,
    yticklabel pos=right,
    ytick={0.02, 0.07, 0.12}, 
    scaled ticks=false,
    yticklabel style={
        anchor=west,
        /pgf/number format/fixed,  
        /pgf/number format/precision=2,  
        /pgf/number format/fixed zerofill,  
    },
},
colormap={mymap}{[1pt]
  rgb(0pt)=(1,1,0.898039215686275);
  rgb(1pt)=(1,0.968627450980392,0.737254901960784);
  rgb(2pt)=(0.996078431372549,0.890196078431372,0.568627450980392);
  rgb(3pt)=(0.996078431372549,0.768627450980392,0.309803921568627);
  rgb(4pt)=(0.996078431372549,0.6,0.16078431372549);
  rgb(5pt)=(0.925490196078431,0.43921568627451,0.0784313725490196);
  rgb(6pt)=(0.8,0.298039215686275,0.00784313725490196);
  rgb(7pt)=(0.6,0.203921568627451,0.0156862745098039);
  rgb(8pt)=(0.4,0.145098039215686,0.0235294117647059)
},
point meta max=0.129336714744568,
point meta min=0.00412825401872396,
tick align=outside,
tick pos=left,
x grid style={darkgray176},
xlabel={Attention Relative Time Step ($t-i$)},
xmin=-0.5, xmax=130.5,
xtick={0, 20, 40, 60, 80, 100, 120},   
xtick style={color=black},
y dir=reverse,
y grid style={darkgray176},
ylabel={Series Index},
ymin=-0.5, ymax=29.5,
ytick={5,10,...,30},   
ytick style={color=black},
width=0.7\textwidth,height=0.3\textwidth
]
\addplot graphics [includegraphics cmd=\pgfimage,xmin=-0.5, xmax=130.5, ymin=29.5, ymax=-0.5] {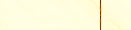};
\end{axis}

\end{tikzpicture}%
    }
    \caption{130 input length.}
    \label{fig:ndt_gma100_len131}
  \end{subfigure}
  \hfill
  \begin{subfigure}[t]{1\textwidth}
    \centering
    \scalebox{0.72}{%
\pgfplotsset{colormap/blackwhite}
\begin{tikzpicture}[trim axis left, trim axis right]

\definecolor{darkgray176}{RGB}{176,176,176}

\begin{axis}[
colorbar,
colorbar style={
    ylabel={Attention weight},
    ylabel style={
        yshift=-.9cm,  
    },
    xshift=.4cm,
    width=.2cm,
    yticklabel pos=right,
    ytick={0.01, 0.015, 0.02}, 
    scaled ticks=false,
    yticklabel style={
        anchor=west,
        /pgf/number format/fixed,  
        /pgf/number format/precision=3,  
        /pgf/number format/fixed zerofill,  
    },
},
colormap={mymap}{[1pt]
  rgb(0pt)=(1,1,0.898039215686275);
  rgb(1pt)=(1,0.968627450980392,0.737254901960784);
  rgb(2pt)=(0.996078431372549,0.890196078431372,0.568627450980392);
  rgb(3pt)=(0.996078431372549,0.768627450980392,0.309803921568627);
  rgb(4pt)=(0.996078431372549,0.6,0.16078431372549);
  rgb(5pt)=(0.925490196078431,0.43921568627451,0.0784313725490196);
  rgb(6pt)=(0.8,0.298039215686275,0.00784313725490196);
  rgb(7pt)=(0.6,0.203921568627451,0.0156862745098039);
  rgb(8pt)=(0.4,0.145098039215686,0.0235294117647059)
},
point meta max=0.0233350079506636,
point meta min=0.00607765931636095,
tick align=outside,
tick pos=left,
x grid style={darkgray176},
xlabel={Attention Relative Time Step ($t-i$)},
xmin=-0.5, xmax=90.5,
xtick={0, 15, 30, 45, 60, 75, 90},   
xtick style={color=black},
y dir=reverse,
y grid style={darkgray176},
ylabel={Series Index},
ymin=-0.5, ymax=29.5,
ytick={5,10,...,30},   
ytick style={color=black},
width=0.7\textwidth,height=0.3\textwidth
]
\addplot graphics [includegraphics cmd=\pgfimage,xmin=-0.5, xmax=90.5, ymin=29.5, ymax=-0.5] {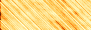};
\end{axis}

\end{tikzpicture}%
    }
    \caption{90 input length.}
    \label{fig:ndt_gma100_len91}
  \end{subfigure}
  \caption{Attention weights at training convergence of TREET optimized by NDG. Each row in the matrices represent a different input sequence and the columns are the weights of past values $i$ from current prediction $t$ (i.e. $i=0$ represent the present prediction $t$). For the GMA(100) process, it can be observed that giving enough time-steps, TREET easily observes at the related time $i=100$ where the information needed to be gathered from, whereas shorter length lead to instability of training.}
\label{fig:ndt_gma100_attn_weights}
\end{figure*}

\subsubsection{AWGN Channel}
\raggedbottom
Consider additive white Gaussian noise (AWGN) channel with i.i.d. noise, 
\begin{align*}
    Z_t &\sim \cN(0, \sigma^2), \\
    Y_t &= X_t + Z_t,\qquad t\in\ZZ,
\end{align*}
$X_t$ is the channel's input sequence, coupled with the average power constraint $\EE[X_t^2]\leq P$. The capacity of this channel is simple for analytical calculation, and is given by the following formula $\text{C}=0.5\log \left(1+ {P}/{\sigma^2}\right)$. Since the process is memoryless, maximized both DI rate and TE (for any $l\geq 0$) coincide with the channel capacity.
We estimated and optimized the TREET according to Algorithm \ref{alg:cap_est_ndt} and compared the model performance with the DI rate estimation and optimization.
Results are presented in Fig. \ref{fig:ndt_awgn_dim} and Fig. \ref{fig:ndt_awgn_db}.
It can be seen that both are estimating the right capacities which are the MI, although their access to multiple past observations.
Moreover, changing the input dimension changes the estimated capacities, as expected. Note that larger dimensions cause error of in estimation and still is an open academic research \cite{song2019understanding, mcallester2020formal, choi2022combating}.
To further analyze the learned distribution, we visualize the optimized NDG mapping in Fig. \ref{fig:ndt_input_output}. It can be seen that the optimized NDG maps the uniform ($U\sim [-1,1]$) inputs into Gaussian samples.
This observation meets our expectations, as the distribution that achieve the channel capacity in AWGN is the Gaussian distribution \cite{CovThom06}.
\raggedbottom

\subsubsection{Gaussian MA(1) Channel}
\raggedbottom
Given a Moving Average (MA) Gaussian noise channel with order 1,
\begin{align*}
    Z_t &= N_t + \alpha N_{t-1},\\
    Y_t &= X_t + Z_t, \qquad t\in\ZZ,
\end{align*}
where $N_t \sim \cN(0,\sigma^2)$ are i.i.d. and $X_t$ is the input to the channel with power constraint $\EE[X_t^2]\leq P$. We apply Algorithm \ref{alg:cap_est_ndt} to both feedforward and feedback settings, comparing with ground truth solutions and the DI-based scheme.
The feedforward capacity can be calculated with the water-filling algorithm \cite{hornik1989multilayer}, whereas the feedback capacity can be calculated by the root of forth order polynomial equation \cite{yang2007feedback}. As seen in Fig. \ref{fig:ndt_gma1_db_fb} and Fig. \ref{fig:ndt_gma1_db_ff}, our method successfully estimated the capacity for a wide range of SNR values, compared to previous method.
\raggedbottom

\subsubsection{Gaussian AR(1) Channel}
\noindent
The case of autoregressive (AR) Gaussian noise channel of order 1 is similar,
\begin{align*}
    Z_t &= N_t + \alpha Z_{t-1},\\
    Y_t &= X_t + Z_t, \qquad t\in\ZZ,
\end{align*}
where $N_t \sim \cN(0,\sigma^2)$ are i.i.d. and $X_t$ is the input to the channel with power constraint $\EE[X_t^2]\leq P$. The capacity is also affected by the existence of feedback. Feedforward capacity can be solved with the water-filling algorithm \cite{hornik1989multilayer} and \cite{sabag2021feedback} prove how to achieve the feedback capacity analytically. Fig. \ref{fig:ndt_gar1_db_fb} and Fig. \ref{fig:ndt_gar1_db_ff} compare the results TREET estimator with the previous one, showcasing it's successful estimation on variated SNR values.
\raggedbottom

\subsection{Capacity Estimation - Memory Analysis}\label{sec:exp_capacity_memory}
Previous experiments showed the capability of correctly estimating TE. Delving deeper into the memory effectiveness of TREET, we tested Gaussian MA channel, as presented before, but with time delay of 100 steps.
\begin{align*}
    Z_t &= N_t + \alpha N_{t-100},\\
    Y_t &= X_t + Z_t, \qquad t\in\ZZ.
\end{align*}
It is notable that to achieve the correct capacity, the information from $t-100$ steps must be an input to the TREET, hence any TE estimation with order $l<100$ will result a false estimation value. The estimation optimization algorithm is tested with shorter input length and longer input length, than the demanded one. Fig. \ref{fig:ndt_gma100_len131} shows the analysis of the attention weights related to the $\hDYXT$ model with $l=130$ input steps (i.e. $\sTE(130)$), at the convergence stages of training, and Fig. \ref{fig:ndt_gma100_len91} considers the same for $l=90$ input steps ($\sTE(90)$). It is evident from Fig. \ref{fig:ndt_gma100_attn_weights} that TREET effectively captures relevant information from long time series when provided with a sufficient input length. However, when the input length is too short, critical information is lost, leading to a complete noisy attention weights.
\raggedbottom

\begin{table}[t]
\centering
\newcommand{\memoryrowlen}{260pt}
\resizebox{
0.48\textwidth
}{!}{%

\begin{tabular}{c
                cc
                cc}
\toprule
  \multirow{3}{*}{\(l\)}
  & \multicolumn{2}{c}{TREET} 
  & \multicolumn{2}{c}{DINE} \\
\cmidrule(lr){2-3}
\cmidrule(lr){4-5}
  & Estimated & Absolute 
  & Estimated & Absolute \\
  & capacity [nat] & error (\%) 
  & capacity [nat] & error (\%) \\
\midrule
60  & 0.19 & 53  & \textbf{0.30} & \textbf{25} \\
70  & 0.29 & 29  & \textbf{0.35} & \textbf{15} \\
80  & 0.28 & 31  & \textbf{0.34} & \textbf{17} \\
90  & 0.29 & 28  & \textbf{0.30} & \textbf{26} \\
100 & \textbf{0.37} & \textbf{9}  
    & 0.36 & 12 \\
110 & \textbf{0.38} & \textbf{7}  
    & 0.33 & 20 \\
120 & \textbf{0.36} & \textbf{11} 
    & 0.33 & 18 \\
130 & \textbf{0.36} & \textbf{11} 
    & 0.34 & 17 \\
140 & \textbf{0.35} & \textbf{14} 
    & 0.26 & 35 \\
\bottomrule
\end{tabular}

}
\caption{Capacity estimation for the GMA(100) channel at 0 SNR with varying memory lengths. The true capacity is $0.405$ [nats]. In this setup, the memory length $l$ for DINE corresponds to the \textit{bptt} length of the LSTM, and for \treetext{}, it represents the sequence input length. DINE estimates the DI rate regardless of input length, whereas \treetext{} estimates the order-$l$ TE. Each value is averaged over ten experiments with different random seeds. Best estimations in \textbf{bold}.}
\label{tab:gma100_results}
\end{table}

Additionally, we compare TREET with DINE for different input lengths on the GMA(100) channel. It is important to note that DINE can theoretically deal with long sequences even if given a shorter backpropagation through time (\textit{bptt}) input length than the process memory due to state propagation. However, as seen in Table \ref{tab:gma100_results}, DINE struggles to estimate the correct capacity under long memory. In contrast, TREET achieves its best estimation for memory lengths larger than 100, with an absolute error rate of less than 14\%. When the TREET memory is shorter than the channel memory, its performance significantly degrades, with absolute error rate no less than 28\%.

\raggedbottom

\subsection{TREET-based Density Estimation}\label{sec:exp_density}
\raggedbottom
In this section, we demonstrate how optimized TREET networks can be used for density estimation - a byproduct of the TREET optimization procedure that requires no additional training. Specifically, TREET enables the estimation of the conditional density function of the observed continuous process $\YY$, namely $P_{Y_t | Y^{t-1}}$, directly from the learned network $\hDYT$. Furthermore, since TREET optimizes both $\hDYT$ and $\hDYXT$, it also provides access to the more informative conditional density $P_{Y_t | Y^{t-1}, X^t}$, enabling flexible estimation depending on the available context.

Given the optimization of $\hDYT$, the network outputs the log–likelihood ratio between the conditional density $P_{Y_t | Y^{t-1}}$ and an explicitly known reference density $P_{\tilde{Y}_t}$. Thus, the true conditional density is recovered via:
\begin{equation}
    P_{Y_t|Y^{t-1}}(y_t | y^{t-1}) \approx \exp{\left(\hDYT(\Dn, g_y)\right)} \cdot \tilde{P}_{Y_t}(y_t).
\end{equation}
This expression defines an unnormalized density function, since the optimized $\hDYT$ approximates the log-likelihood ratio (up to an additive constant) between the true conditional density and the reference density, as guaranteed by the DV representation. Consequently, the probability given by this equation is not directly normalized. To obtain a properly normalized conditional density, the continuous domain of $Y_t$ is discretized into a sufficiently fine grid, evaluate the unnormalized density at each grid point, and numerically normalize by summing over all points. In practice, this numerical integration approach provides a valid approximation of the conditional density.

Our method builds upon previous work \cite{park2024deep}, which addressed non–time-related density estimation, and \cite{aharoni2022density}, who proposed a time-related approach based on the DINE framework. While DINE is designed to capture limiting (stationary) densities, TREET leverages a fixed memory length, aligning naturally with memory-aware estimation.

The hidden and observed processes are modeled as
\begin{align*}\label{eq:gaussian_hmm}
    X_t &= \alpha X_{t-1} + \beta X_{t-k} + W_t, \\
    Y_t &= \gamma X_t + V_t, \numberthis
\end{align*}
where $W_t,V_t$ are i.i.d. innovations (either Gaussian, zero mean and $\sigma^2=0.5$, or uniform on $[-1,1]$). The process is stationary (and ergodic) iff every root $z$ of the characteristic polynomial $1-\alpha z-\beta z^{k}=0$ lies outside the unit circle, $|z|>1$ \cite{brockwell1991time}. We conduct two experiments - the first is classic HMM without any state delay ($k=0$), with parameters $\alpha=0.9$, $\beta=0$, $\gamma=0.5$, $\sigma_W^2=\sigma_V^2=0.5$, 
and fix the memory length of all models to $l=3$. TREET employs this $l=3$ memory parameter for TE estimation, and DINE’s \textit{bptt} is likewise truncated to three steps.
The second experiment is memory analysis by applying state-delay, with $\beta\neq0$ and variate delay $k$. Set $\alpha=0.001$, $\beta=0.9$, $\gamma=0.9$, $\sigma_W^2=\sigma_V^2=0.5$, and TREET’s memory $l=k+5$ and compared all but KDE due to computational complexity of the algorithm for larger $k$s.

For the classic HMM, TREET compared against several baselines: DINE; mixture density networks (MDN)\cite{bishop1994mixture}, which output a Gaussian mixture model for conditional densities; conditional KDE\cite{o2016fast}, which performs non-parametric smoothing over the conditioning variables; and the Kalman filter\cite{kalman1960new}, the optimal solution in the Gaussian HMM (without delay) setting. 
To evaluate TREET’s density-estimation performance, we compare the estimated conditional density of observations $P_{Y_t|Y^{t-1}}$ to the analytical conditional density $P_{Y_t|X_t}$, which assumes full access to the latent state. As metrics, KL divergence  is computed and the total variation (TV) distance between each model’s estimate and the analytical density. 

\begin{table}[t]
    \centering
    \newcommand{\pderowlen}{170pt}
    \resizebox{0.33\textwidth}{!}{



    \begin{tabular}{lcccc}
        \toprule
        \multirow{2}{*}{Model} & \multicolumn{2}{c}{Gaussian} & \multicolumn{2}{c}{Uniform} \\ 
        \cmidrule(lr){2-3} \cmidrule(lr){4-5}
        & KLD & TV & KLD & TV \\ 

        \midrule
        Kalman &  1.076  &  0.467  &  1.304  &  0.408  \\ 
        \midrule
        KDE &  1.135  &  0.608  &  0.847  &  0.417  \\ 
        \midrule
        MDN &  1.098  &  0.632  &  0.667  &  0.460  \\ 
        \midrule
        DINE & \textbf{0.795} & \underline{0.525} & \textbf{0.475} & \textbf{  0.291 } \\ 
        \midrule
        TREET & \underline{0.797} & \textbf{0.524} & \underline{ 0.482 } & \underline{ 0.296 } \\
        
        \bottomrule
    \end{tabular}%


    }
    \caption{Comparison of density estimation models on the HMM process without state delay ($\beta=0$), with Gaussian and uniform noise. Results show KL divergence and TV distance relative to the analytical conditional density, averaged over ten trials. Best in \textbf{bold}; second-best \underline{underlined}}
    \label{tab:pde_hmm}
\end{table}

\begin{table}[t]
    \centering
    \newcommand{\pderowlen}{170pt}
    \resizebox{0.49\textwidth}{!}{

\begin{tabular}{c
                cc
                cc
                cc
                cc}
\toprule
\multirow{2}{*}{$k$}
  & \multicolumn{2}{c}{TREET} 
  & \multicolumn{2}{c}{DINE} 
  & \multicolumn{2}{c}{MDN} 
  & \multicolumn{2}{c}{Kalman} \\
\cmidrule(lr){2-3}
\cmidrule(lr){4-5}
\cmidrule(lr){6-7}
\cmidrule(lr){8-9}
  & KL & TV 
  & KL & TV 
  & KL & TV 
  & KL & TV \\
\midrule
2  &  \underline{0.933}      &  \underline{0.569}         &  0.934  & 0.570          & 1.460 & 0.708        & \textbf{0.636} & \textbf{0.405} \\
5  & \underline{1.036}      & 0.601         & \textbf{0.875}  & \textbf{0.556}          & 1.454 & 0.707        & 1.284 & \underline{0.599} \\
10 & \textbf{0.984}      & \textbf{0.585}         & 1.299  & 0.662          & 1.450 & 0.706        & \underline{1.239} & \underline{0.603} \\
15 & \textbf{0.992}      & \textbf{0.587}         & 1.441  & 0.704          & 1.448 & 0.707        & \underline{1.232} & \underline{0.602} \\
25 & \textbf{0.993}      & \textbf{0.587}         & 1.451  & 0.706          & 1.443 & 0.704        & \underline{1.196} & \underline{0.598} \\
50 & \textbf{0.993}      & \textbf{0.589}         & 1.452  & 0.707          & 1.453 & 0.707        & \underline{1.193} & \underline{0.597} \\
\bottomrule
\end{tabular}

    }
    \caption{Memory-delay analysis of density estimation on the HMM process ($\beta\neq0$) with varying state delay $k$. TREET and DINE are compared in terms of mean KL divergence and TV distance relative to the analytical density, averaged over ten trials. Best in \textbf{bold}; second-best \underline{underlined}}
    \label{tab:pde_hmm_memory}
\end{table}

Table \ref{tab:pde_hmm} reports these distances for the classic HMM experiment averaged over ten trials. Under both Gaussian and uniform noise, TREET matches or exceeds the performance of DINE, mixture density networks, conditional kernel density estimation, and the Kalman filter.
Table \ref{tab:pde_hmm_memory} summarizes the memory-delay analysis. For each state delay $k\in\{2,5,10,15,25,50\}$ we set the memory for each algorithm to $l=k+5$, e.g. TE parameter for TREET and \textit{bptt} for DINE.
and report mean KL divergence and TV distance over ten trials. 
TREET matches or improves upon DINE’s KL and TV scores for every state delay, and begins to pull away from both MDN and the Kalman filter once $k \geq 10$, demonstrating its ability to capture long-range dependencies. At very short delays, however, the Kalman filter and DINE remains competitive - reflecting their explicit state-update schemes suited to near-Markovian dynamics.

Overall, these two experiments confirm that TREET provides accurate, memory-aware conditional density estimates with no additional training beyond TE optimization, outperforming existing methods even as the required memory length grows.

\subsection{Features Analysis in Physiological Data}\label{sec:exp_real_world}
Motivated by the results in \cite[Chapter 7.1]{Bossomaier2016AnIT}, we tested the TREET on the Apnea dataset from  Santa Fe Time Series Competition\footnote{\underline{\url{https://physionet.org/content/santa-fe/1.0.0/}}} \cite{rigney1993apnea, goldberger2000apnea}. This is a multivariate data that have been recorded from a diseased patient of Apnea in a sleep laboratory, which is  a condition characterized by brief, involuntary pauses in breathing, particularly during sleep. Each sample consists of three different variables from a specific time, with a sample rate of 2 Hz. The three features are heart rate, chest volume (which is the respiration force) and blood oxygen concentration. A sampled sequence of heart rate and breath rate (chest volume) is presented in Fig. \ref{fig:apnea_data}.

\begin{figure*}[t]
    \begin{subfigure}[t]{.5\textwidth}
        \centering
        \scalebox{.8}{%

\begin{tikzpicture}[trim axis left, trim axis right]

\definecolor{darkgray176}{RGB}{176,176,176}

\begin{groupplot}[group style={group size=1 by 2}]
\nextgroupplot[
scaled x ticks=manual:{}{\pgfmathparse{#1}},
tick align=outside,
tick pos=left,
x grid style={darkgray176},
xlabel={Time},
xmajorgrids,
xmin=190.05, xmax=408.95,
xtick style={color=black},
xticklabels={},
y grid style={darkgray176},
ylabel={Amplitude},
ymajorgrids,
ymin=56.5845, ymax=91.0255,
ytick style={color=black},
width=0.6\textwidth,
height=0.3\textwidth,
legend style={nodes={scale=0.8, transform shape}, at={(0.02,0.02)}, anchor=south west, font=\small, cells={anchor=west}},
width=1\textwidth,height=0.5\textwidth
]
\addplot [line width=0.3pt, red]
table {%
200 79.48
201 80.04
202 79.96
203 80.34
204 80.82
205 81.04
206 81.08
207 80.9
208 80.3
209 79.21
210 78.53
211 78.85
212 79.91
213 81.37
214 82.59
215 83.68
216 83.83
217 82.59
218 81
219 78.97
220 78.79
221 81.14
222 82.19
223 81.73
224 81.23
225 79.95
226 79.29
227 79.93
228 80.44
229 81.13
230 82.06
231 82.92
232 83.03
233 82.58
234 82.42
235 82.08
236 81.23
237 79.62
238 79.65
239 80.96
240 80.69
241 81.42
242 82.68
243 82.12
244 82.3
245 82.28
246 80.21
247 79.52
248 79.89
249 79.32
250 81.1
251 83.42
252 83.46
253 83.33
254 82.19
255 80.13
256 78.07
257 78.19
258 80.71
259 81.74
260 81.03
261 80.02
262 77.44
263 74.45
264 72.47
265 74.57
266 78.54
267 79.37
268 80.02
269 80.4
270 78.92
271 76.71
272 75.63
273 77.26
274 79.34
275 80.15
276 81.71
277 83.03
278 83.55
279 84.81
280 86.34
281 86.86
282 86.41
283 85.17
284 83.82
285 83.21
286 83.43
287 83.62
288 83.02
289 81.84
290 80
291 78.47
292 77.47
293 76.34
294 75.86
295 77.46
296 78.81
297 77.76
298 77.76
299 77.76
300 77.76
301 77.82
302 80.56
303 82.71
304 84.35
305 85.51
306 86.92
307 88.76
308 89.46
309 88.7
310 87.75
311 85.85
312 84.26
313 83.97
314 84.56
315 84.45
316 82.38
317 77.86
318 74.63
319 74.63
320 75.38
321 75.59
322 74.28
323 73.39
324 73.1
325 72.36
326 73.69
327 77.7
328 77.82
329 77.02
330 80.08
331 82.82
332 83.28
333 82.4
334 81.06
335 80.68
336 81.53
337 81.49
338 81.75
339 83.02
340 84.42
341 83.85
342 81.65
343 79.88
344 80.77
345 83.05
346 83.19
347 81.53
348 79.1
349 76.51
350 74.44
351 73.58
352 72.53
353 72.01
354 72.54
355 72.54
356 72.54
357 65.68
358 65.68
359 58.15
360 58.86
361 59.92
362 61.97
363 65.06
364 67.39
365 68.46
366 70.19
367 72.85
368 77.38
369 81.16
370 81.96
371 82.56
372 82.48
373 82.58
374 83.91
375 85.22
376 84.23
377 81.7
378 80.65
379 81.18
380 81.65
381 79.25
382 77.16
383 77.52
384 76.89
385 76.33
386 76.72
387 79.02
388 81.32
389 81.52
390 81.1
391 79.39
392 76.75
393 73.78
394 71.72
395 70.36
396 71.27
397 75.01
398 74.53
399 71.78
};
\addlegendentry{Heart Rate}

\nextgroupplot[
tick align=outside,
tick pos=left,
x grid style={darkgray176},
xlabel={Time},
xmajorgrids,
xmin=190.05, xmax=408.95,
xtick style={color=black},
y grid style={darkgray176},
ylabel={Amplitude},
ymajorgrids,
ymin=-5881.8, ymax=18379.8,
ytick style={color=black},
width=0.6\textwidth,
height=0.3\textwidth,
legend style={nodes={scale=0.8, transform shape}, at={(0.02,0.98)}, anchor=north west, font=\small, cells={anchor=west}},
width=1\textwidth,height=0.5\textwidth
]
\addplot [line width=0.3pt, blue]
table {%
200 6469
201 4744
202 2798
203 3532
204 929
205 319
206 2135
207 10123
208 8452
209 6502
210 6023
211 6528
212 1523
213 714
214 2716
215 10075
216 11764
217 8221
218 6514
219 6771
220 4964
221 2251
222 730
223 7984
224 6165
225 6456
226 7243
227 7266
228 7342
229 6209
230 5940
231 2776
232 2661
233 1574
234 2206
235 6700
236 7591
237 7983
238 7246
239 4121
240 3704
241 4094
242 2434
243 1607
244 2576
245 9300
246 8711
247 7306
248 7167
249 7658
250 2161
251 1058
252 2548
253 9748
254 9188
255 7259
256 6822
257 6274
258 3866
259 2819
260 1311
261 3382
262 9756
263 7225
264 6331
265 1927
266 -372
267 1250
268 11094
269 9835
270 9296
271 7600
272 5349
273 -289
274 -1731
275 2284
276 11996
277 9010
278 4855
279 5455
280 5807
281 9374
282 7146
283 5098
284 2347
285 -807
286 4
287 10184
288 10563
289 6505
290 5510
291 4645
292 6019
293 6094
294 6359
295 5268
296 4075
297 4216
298 5312
299 4823
300 1559
301 -255
302 3596
303 7769
304 7744
305 6846
306 6647
307 8410
308 6687
309 8016
310 5300
311 4236
312 4254
313 5112
314 5119
315 3373
316 3524
317 4173
318 3716
319 4600
320 4514
321 4904
322 5425
323 5568
324 5522
325 5509
326 5520
327 5390
328 -2623
329 -920
330 4699
331 13317
332 9582
333 4055
334 -1213
335 5859
336 7402
337 7870
338 7526
339 13119
340 9718
341 8823
342 6601
343 1835
344 -3368
345 -1321
346 10068
347 8437
348 5559
349 4843
350 5113
351 5368
352 5822
353 6495
354 4434
355 3823
356 3481
357 3818
358 4195
359 4494
360 4858
361 4974
362 5465
363 5389
364 5691
365 2400
366 -3048
367 444
368 3715
369 17277
370 11695
371 5905
372 5336
373 5305
374 6888
375 7643
376 7193
377 3355
378 -4779
379 -2381
380 14135
381 10468
382 6818
383 3846
384 4410
385 4878
386 5011
387 6077
388 6262
389 8959
390 7534
391 4819
392 5655
393 5191
394 3354
395 4286
396 1007
397 -490
398 9306
399 8440
};
\addlegendentry{Breath Rate}
\end{groupplot}

\end{tikzpicture}

        }
        \caption{Sampled sequence of Apnea dataset.}
        \label{fig:apnea_data}
    \end{subfigure}
    \hfill
    \begin{subfigure}[t]{.5\textwidth}
        \centering
        \scalebox{.8}{%
\begin{tikzpicture}[trim axis left, trim axis right]

\definecolor{darkgray176}{RGB}{176,176,176}

\begin{axis}[
tick align=outside,
tick pos=left,
x grid style={darkgray176},
xlabel={Label Length \(\displaystyle [l]\)},
xmajorgrids,
xmin=2.4, xmax=15.6,
xtick style={color=black},
y grid style={darkgray176},
ylabel={\(\displaystyle TE_{X\to Y}(k,l=2)\)},
ymajorgrids,
ymin=-0.003043, ymax=0.063903,
ytick style={color=black},
width=0.6\textwidth,
height=0.4\textwidth,
legend style={nodes={scale=0.8, transform shape}, at={(0.98,0.98)}, anchor=north east, font=\small, cells={anchor=west}},
width=1\textwidth,height=0.9\textwidth
]
\addplot [draw=black, mark=o, only marks]
table{%
x  y
3 0.04825
4 0.05115
5 0.0571
6 0.06086
7 0.05689
8 0.04878
9 0.04427
10 0.04578
11 0.0436
12 0.03619
13 0.03913
14 0.03845
15 0.03835
};
\addlegendentry{Breath $\rightarrow$ Heart}

\addplot [draw=black, fill=black, mark=x, only marks]
table{%
x  y
3 0
4 0
5 0
6 0
7 0
8 0.005303
9 0
10 0.01177
11 0
12 0
13 0
14 0
15 0
};
\addlegendentry{Heart $\rightarrow$ Breath}

\end{axis}

\end{tikzpicture}
        }
        \caption{TE estimation for variable length history of $Y$.}
        \label{fig:apnea_fixed_xlen}
    \end{subfigure}
    \caption{Transfer Entropy estimation on physiological data. The Apnea dataset consists of heart rate and breathing rate measurements. The information flow for patient diagnosis is diagnosed. Patients with Apnea experience breathing cessation, leading to alterations in heart rate during sleep.}
    \label{fig:apnea}
\end{figure*}
\raggedbottom

Determining the interaction between different physiological features is crucial for diagnosing diseases by revealing causal connections within the human body, enabling targeted diagnostics and personalized treatments according to identified risk factors.
Therefore, TREET was applied to determine the magnitude and direction of information transfer in the given setting.
Both heart rate and breath features were used to compare the TE measurements $\mathsf{TE}_{\mathsf{Breath} \to \mathsf{Heart}}(k, 2)$ and $\mathsf{TE}_{\mathsf{Heart} \to \mathsf{Breath}}(k, 2)$ for variable length $k$ that is the $Y$ process's history observations length. The results are presented in Fig. \ref{fig:apnea_fixed_xlen} and are aligned with the results of \cite{Bossomaier2016AnIT} (for $k,l=2$) and extend it with tests on variate history length of $Y$ process.

Notably, for every considered $k$, the TE from the breath process to the heart process is consistently higher, aligning with the diagnosis of Apnea disorder (abrupt cessation of breathing during sleep).
In terms of information flow, our results indicate that the breathing process transfers more information regarding the behavior of the heart rate process, than the opposite direction, since the value of the estimated TE is consistently higher in comparison, whereas in information theory it essentially reflect that the $X$ process has more to reveal about $Y$'s current value than $Y$'s past itself.

Furthermore, in the influence direction $\mathsf{TE}_{\mathsf{Breath} \to \mathsf{Heart}}$, 
increasing the number of visible heart rate history samples decreases the information transfer, since a longer history of heart rate provides more insight into future outcomes than the instantaneous breath value.
However, $\mathsf{TE}_{\mathsf{Heart} \to \mathsf{Breath}}$ shows that for variate $k$ values, the majority of TE results indicate that the heart process barely affect the breathing process in Apnea patients. 

The conclusion of this experiment provides valuable insights and validates established scientific knowledge. While it is generally understood that an increase in heart rate corresponds with an increase in breathing rate, patients with Apnea exhibit a distinct phenomenon: a sudden cessation of muscle activity during inhalation (i.e., breathing cessation) significantly affects the heart rate, contrary to the behavior observed in healthy individuals. This experiment corroborates the findings presented in \cite{Bossomaier2016AnIT, tsur2024infomat} and further introduces a novel insight - namely, that the estimated TE decreases with increasing context (history length) of the process $Y$. This observation aligns with the theoretical expectation that incorporating a longer historical context of $Y$ reduces uncertainty, thus lowering the resulting TE values.
\raggedbottom


\section{Conclusions and Future Work}\label{sec:conclusions_future_work}
\noindent This work presented an attention-based architecture for estimating TE in ergodic and stationary processes. We developed a DV-based neural TE estimator, established its consistency, and introduced a novel modified attention mechanism tailored for the task. Our extended TE benchmark demonstrates its superior performance relative to existing methods. 
Furthermore, we designed an optimizer for the estimated TE, which was subsequently applied to channel capacity estimation, and proved to perform greatly for higher order TE estimation. 
In addition, showcased its inherent probability density estimation functionality, which emerges directly from the TE estimation training and finally evaluated TREET's capability in causal feature analysis on the Apnea dataset.

With the increasing popularity of sequential data in contemporary machine learning, TREET will be leveraged for information theoretic analysis and architecture design through the lens of causal information transfer. Potential applications include enhancing predictive probabilistic models, refining feature selection processes, reconstructing complex networks, improving anomaly detection capabilities, and optimizing decision-making in dynamic environments. Moreover, building on the work in \cite{tsur2024rate}, the TREET optimization scheme will be extended to sequential data compression tasks.

\raggedbottom
\section{Appendix}\label{sec:appendix}
\subsection{Proof of lemma 1
}\label{sec:proof_lemma_te_di_rate_relation_markov}
Let $\XX$, $\YY$ be two jointly stationary processes that pose the markov property with $l\in\NN$ and $m\geq l$, then
\begingroup
\allowdisplaybreaks
\begin{align*}\label{eq:te_equals_di_proof_1}
    &\sTE(l)\\
    &= \sI\left( X^{l};Y_{l} | Y^{l-1}\right)\\ 
    &\stackrel{(a)}{=} \sI \left(X^{l}_{l-m}; Y_{l} | Y^{l-1}_{l-m} \right)\\
    &\stackrel{(b)}{=} \sI\left( X^m;Y_m | Y^{m-1}\right)\\ 
    &= \sTE(m),\numberthis
\end{align*}
\endgroup
where (a) is true from the markovity, and (b) transition is index shift which valid due to process stationarity.
Observing the DI rate,
{\allowdisplaybreaks
\begin{align*}\label{eq:te_equals_di_proof_2}
    &\sI(\left( \XX \to \YY \right)\\
    &\stackrel{(a)}{=} \lim_{n \to \infty} \frac{1}{n}\sum_{i=1}^{n} \sI\left(X^i; Y_i | Y^{i-1} \right)\\
    &= \lim_{n \to \infty} \frac{1}{n}\sum_{i=1}^{n}\left[ \sh \left(Y_{i} |Y^{i -1 }\right)- \sh \left(Y_{i} |X^{i}, Y^{i -1 }\right)\right]\\
    &\stackrel{(b)}{=} \lim_{n \to \infty} \sh \left(Y_{n} |Y^{n -1 }\right) - \sh \left(Y_{n} |X^{n}, Y^{n -1 }\right)\\
    &= \lim_{n \to \infty} \sI\left( X^{n};Y_{n} | Y^{n - 1}\right)\\
    &\stackrel{(c)}{=} \lim_{n \to \infty} \sTE(n)\\
    &\stackrel{(d)}{=} \sTE(m), \numberthis
\end{align*}
}where the limit in (a) exists whenever the joint process is stationary, transition (b) is valid because the limit exists for each series of conditional entropies, and since conditioning reduces entropy, the limit for each normalized sum of series is $\lim_{n \to \infty} \sh \left(Y_{n} |Y^{n -1 }\right), \lim_{n \to \infty} \sh \left(Y_{n} |X^{n}, Y^{n -1 }\right)$, respectively \cite[Theorem 4.2.1]{CovThom06}. Since the limit exists for the conditional MI, transition (c) is valid by definition of TE, and the TE with limit of parameter $m$ exists, and (d) is from \eqref{eq:te_equals_di_proof_1} for $n\geq m$. Concluding the proof. $\square$

\subsection{Proof of theorem 3
}\label{sec:proof_tene_consistency}
\raggedbottom
The proof following the steps of representation step - represents TE as a subtraction of two DV potentials, 
estimation step - proves that the DV potentials is achievable by empirical mean of a given set of samples, 
and approximation step - shows that the estimator converges to TE with the corresponding memory parameter $l$.
Thus concluding that our estimator is a consistent estimator for the TE.

Let $\{X_i,Y_i\}_{i\in\ZZ}$ be the two values of a process $\XX, \YY$ respectively, and $\PP$ be the stationary ergodic measure over $\sigma(\XX,\YY)$. Define $P_{X^n,Y^n}:=\PP|_{\sigma(X^n,Y^n)}$ as the $n$-coordinate projection of $\PP$, where $\sigma(X^n,Y^n)$ is the $\sigma$-algebra generated by $(X^n,Y^n)$. Let $\Dn=(X^n,Y^n)\sim P_{X^n,Y^n}$. Let $\tilde{Y}\sim {Q}_{Y}$ be an absolutely continuous PDF, independent of $\{(X_i,Y_i)\}_{i\in\ZZ}$ and its distribution noted as $\tilde{P}_{Y}$. The proof is divided to three steps - variational representation, estimation from samples and functional approximation.

\subsubsection{Representation of TE}\label{subsec:representation_te}
\noindent
In order to write TE as a difference between two KL divergences, recall Lemma \ref{lemma:te_dkl} - let,
\allowdisplaybreaks
\begin{subequations}\label{eq:dy_dyxt_def_recall}
        \begin{align}
            &\DYT := \DKL \left(P_{Y_l|Y^{l-1}} \| \tilde{P}_{Y_l} \Big{|} P_{Y^{l-1}} \right),\\
            &\DYXT := \DKL \left(P_{Y_l|Y^{l-1}X^l} \| \tilde{P}_{Y_l} \Big{|} P_{Y^{l-1}X^{l}} \right).
        \end{align}
    \end{subequations}
    Then
    \begin{equation}\label{eq:te_dkl_recall}
        \sTE(l) = \DYXT - \DYT.
    \end{equation}
This lemma is proved in Appendix \ref{subsec:te_dkl_proof}.
Estimating the KL divergence is applicable with the DV representation \ref{theorem:dv_representation}
\begin{subequations}
    \begin{align*}
        \DYT =& \sup_{f_y:\Omega_\cY\to\RR} \EE \left[ f_y \left(Y^{l}\right)\right] \\
        &  -\log \EE \left[ e^{f_y \left(Y^{l-1}, \tilde{Y}_{l}\right)}\right], \numberthis \label{eq:dy_dv_repres}
    \end{align*}
    where $\Omega_\cY = \cY^{l}$. For the other term, the DV representation is,
    \begin{align*}
        \DYXT =& \sup_{f_{xy}:\Omega_{\cX \times \cY} \to \RR} \EE \left[ f_{xy} \left(X^{l},Y^{l}\right)\right] \\
        & -\log \EE \left[ e^{f_{xy} \left(X^{l},Y^{l-1}, \tilde{Y}_l\right)}\right],\numberthis \label{eq:dyx_dv_repres}
    \end{align*}
\end{subequations}
where $\Omega_{\cX\times\cY} = \cX^{l} \times \cY^{l}$.
The next section refers to \eqref{eq:dy_dv_repres} but the claims are the same for \eqref{eq:dyx_dv_repres}.

\subsubsection{Estimation}
\noindent
By the DV representation, the supremum in \eqref{eq:dy_dv_repres} achieved for
\begin{align*}\label{eq:f_star_y}
    f_{y,l}^{\star}&:=\log \left( \frac{dP_{Y^l}}{d(P_{Y^{l-1}}\otimes \tilde{P}_{Y_l})}\right)\\
    &=\log p_{Y_l | Y^{l-1}}-\log \tilde{p}_{Y_l}, \numberthis
\end{align*}
where the last equality holds due to $P_{Y^{l}} \ll P_{Y^{l-1}} \otimes \tilde{P}_{Y_l}$, both measures have Lebesgue densities. 
Although it is not mandatory to select the reference measurement as uniform, choosing a uniform reference measurement can result in a constant that can be neutralized to obtain the likelihood function of $Y_l|Y^{l-1}$. This approach allows for simplification and facilitates the estimation process.
Empirical means can estimate the expectations in \eqref{eq:dy_dv_repres}, while applying the generalized Birkhoff theorem \cite{breiman1957individual}, stated next:

\begin{theorem}[The generalized Birkhoff theorem]\label{theorem:birkhoff}
    Let $T$ be a metrically transitive one-to-one measure preserving transformation of the probability space $(\Omega, \mathcal{F} , \PP)$ onto itself. Let $g_0(\omega), g_1(\omega), \ldots$ be a sequence of measurable functions on $\Omega$ converging a.s. to the function $g(\omega)$ such that $\EE[\sup_i |g_i|] \leq \infty$. Then,
    \begin{equation}\label{eq:birkhoff}
        \frac{1}{n}\sum_{i=1}^{n}g_i(T^i\omega) \stackrel{n\to\infty}{\longrightarrow{}} \EE[g],\quad \PP-a.s.
    \end{equation}
\end{theorem}
By applying Theorem \ref{theorem:birkhoff} and for any $\epsilon>0$ and sufficiently large $n$, we have
\begin{subequations}\label{eq:bounds_for_fys}
    \begin{align*}
        &\left| \EE
            \left[f_{y,l}^\star
                \left(Y^l
                \right)
            \right] -\frac{1}{n}\sum_{i=0}^{n-1}f_{y,l}^\star
            \left(Y_{i}^{i+l}
            \right)
        \right| < \frac{\epsilon}{8},\numberthis \label{eq:fy_empirical}\\
        &\left| \log 
            \left( \EE
                \left[ e^{f_{y,l}^\star
                        \left( Y^{l-1},\tilde{Y}_l
                        \right)}
                \right]
            \right) \right.\\
            &  \quad \left. -\log
            \left(\frac{1}{n}\sum_{i=0}^{n-1}e^{f_{y,l}^\star
                \left(Y_{i}^{i+l-1},\tilde{Y}_{i+l}
                \right)}
            \right)
        \right| < \frac{\epsilon}{8},\numberthis \label{eq:fy_tilde_empirical}
    \end{align*}
\end{subequations}
where $\{f_{y,l}^\star\}$ is the function of $l$ time steps, that achieves the supremum of $\DKL \left(P_{Y_l|Y^{l-1}} \| \tilde{P}_{Y_l} \Big{|} P_{Y^{l-1}} \right)$. Convergence achieved from the generalized Brikhoff theorem, where the series of functions is the fixed function $f_{y,l}^\star$,

\begin{subequations}
    \begin{align}
        &\EE_n
            \left[
                f_{y,l}^\star
                \left(
                    Y^l
                \right)
            \right]:=\frac{1}{n}\sum_{i=0}^{n-1}f_{y,l}^\star
            \left(Y_{i}^{i+l}
            \right), \label{eq:en_fy}\\
        &\EE_n
            \left[
                e^{f_{y,l}^\star
                \left(
                    Y^{l-1}, \tilde{Y}_l
                \right)}
            \right]:=\frac{1}{n}\sum_{i=0}^{n-1}e^{f_{y,l}^\star
            \left(Y_{i}^{i+l-1},\tilde{Y}_{i+l}
            \right)}. \label{eq:en_fy_tilde}
    \end{align}
\end{subequations}

\subsubsection{Approximation}
\noindent
Last step is to approximate the functional space with the space of transformers. Recall that set of causal transformer architectures $\GTFC^Y:=\GTFC^{(d_y,1,l,v_y)}, \GTFC^{XY}:=\GTFC^{(d_x+d_y,1,l,v_{xy})}$ for given $l,d_y,d_x,v_y,v_{xy} \in \NN$. Define 
\begin{align*}
    &\hDYT(\Dn):=\sup_{g_y\in \GTF^Y}\frac{1}{n}\sum_{i=0}^{n-1}g_y\left(Y_{i}^{i+l}\right)\\
    &  \qquad \qquad \quad -\log\left( \frac{1}{n}\sum_{i=0}^{n-1}e^{g_y\left(Y_{i}^{i+l-1},\tilde{Y}_{i+l}\right)}\right)\numberthis,
\end{align*}
where the DV functions are transformers $g_y\in\GTFC^{Y}, g_{xy}\in\GTFC^{XY}$. 
We want to prove that for a given $\epsilon>0$
\begin{equation}
    \left| \hDYT(\Dn)-\DYT\right|\leq \frac{\epsilon}{2}.
\end{equation}
From Theorem \ref{theorem:dv_representation}, we obtain
\begin{subequations}
    \begin{align*}
        &\EE 
            \left[ f_{y,l}^\star
                \left(Y^l 
                \right) 
            \right]=\DYT,\numberthis\\
        &\EE
            \left[f_{y,l}^\star 
                \left(Y^{l-1},\tilde{Y}_l
                \right) 
            \right]=1.\numberthis
    \end{align*}
\end{subequations}
Thus, we seek to bound the following subtraction $\left| \hDYT(\Dn) - \EE\left[ f_{y,l}^\star\left(Y_{i}^{i+l}\right)\right]\right|$. By the given identity $\log(x)\leq x-1, \forall x\in\RR_{\geq 0}$,
\begin{align*}
        &\left|
            \hDYT(\Dn)-\EE\left[
            f_{y,l}^\star\left(Y^l
                \right)
            \right]
        \right|\\
        &=\left|
            -\EE\left[
            f_{y,l}^\star\left(Y^l
                \right)
            \right] + 
            \sup_{g_y\in \GTF^Y} \left\{\frac{1}{n}\sum_{i=0}^{n-1}g_y \left( Y_{i}^{i+l}\right) \right. \right. \\
            & \quad  \left. \left. -log \left(
                \frac{1}{n}\sum_{i=0}^{n-1}e^{g_y \left(Y_{i}^{i+l-1}, \tilde{Y}_{i+l} \right)}
            \right)\right\}
        \right|\\
        &\leq\left|
            1
            -\EE\left[
            f_{y,l}^\star\left(Y^l
                \right)
            \right] + 
            \sup_{g_y\in \GTF^Y} \left\{\frac{1}{n}\sum_{i=0}^{n-1}g_y \left( Y_{i}^{i+l}\right) \right. \right. \\
            & \quad \left. \left. - \left(
                \frac{1}{n}\sum_{i=0}^{n-1}e^{g_y \left(Y_{i}^{i+l-1}, \tilde{Y}_{i+l}\right)}
            \right)\right\}
        \right|\\
        &\leq\left|
            +\EE\left[f_{y,l}^\star 
                    \left(Y^{l-1},\tilde{Y}_l
                    \right) 
                \right]
            -\EE\left[
            f_{y,l}^\star\left(Y^l
                \right)
            \right] \right. \\
            & \quad \left. + \sup_{g_y\in \GTF^Y} \left\{ \frac{1}{n}\sum_{i=0}^{n-1}g_y \left( Y_{i}^{i+l}\right) \right. \right. \\
            & \quad \left. \left. - \left(
                \frac{1}{n}\sum_{i=0}^{n-1}e^{g_y \left(Y_{i}^{i+l-1}, \tilde{Y}_{i+l}\right)}
            \right)\right\} 
        \right|.
        \numberthis\label{eq:dv_bound}
\end{align*}
Due to \eqref{eq:bounds_for_fys}, there exists $N\in\NN$ such that $\forall n > \NN$
\begin{subequations} \label{eq:empirical_mean_bound}
    \begin{align*}
        &\left| \EE
            \left[f_{y,l}^\star
                \left(Y^{l}
                \right)
            \right] -\EE_n
            \left[
                f_{y,l}^\star
                \left(
                    Y^{l}
                \right)
            \right]
        \right| < \frac{\epsilon}{8},\numberthis \\
        &\left| 
            \left( \EE
                \left[ e^{f_{y,l}^\star
                        \left( Y^{l-1},\tilde{Y}_l
                        \right)}
                \right]
            \right) -\EE_n
            \left[
                e^{f_{y,l}^\star
                \left(
                    Y^{l-1}, \tilde{Y}_l
                \right)}
            \right]
        \right| < \frac{\epsilon}{8}.
        \numberthis
    \end{align*}
\end{subequations}
Plugging \eqref{eq:empirical_mean_bound} to \eqref{eq:dv_bound} gives
\begin{align*}
    &\left|
        \hDYT(\Dn)-\DYT
    \right|\\
    &\leq  \frac{\epsilon}{4} + \Bigg{|} - \EE_n \left[
            e^{f_{y,l}^\star
            (
                Y^{l-1}, \tilde{Y}_l
            )}
        \right] - 
        \EE_n \left[
            f_{y,l}^\star
            \left(
                Y^l
            \right)
        \right]  \\
    & \quad +\sup_{g_y\in \GTF^Y} 
    \left\{ \frac{1}{n}\sum_{i=0}^{n-1}g_y 
        \left( Y_{i}^{i+l}\right) \right. \\ 
        & \quad \left. -\left(
            \frac{1}{n}\sum_{i=0}^{n-1}e^{g_y 
            \left(Y_{i}^{i+l-1}, \tilde{Y}_{i+l}\right)}
        \right)
    \right\} 
    \Bigg{|}.
    \numberthis \label{eq:dv_bound_22}
\end{align*}
Since the empirical mean of $f_{y,l}^\star$ is converging to the expected mean, it is uniformly bounded by some $M\in\RR_{\geq0}$. Since the exponent function is Lipschitz continuous with Lipschitz constant $\exp{[M]}$ on the interval $(-\infty,M]$, we obtain
\allowdisplaybreaks
\begin{align*}
    &\frac{1}{n}\sum_{i=1}^n e^{f_{y,l}^\star \left(\tilde{Y}_{i+l},Y_{i}^{i+l-1}\right)} - e^{g_y\left(\tilde{Y}_l,Y_{i}^{i+l-1}\right)} \\
    &\leq e^M\frac{1}{n}\sum_{i=1}^{n}\Big{|} f_{y,l}^\star \left(\tilde{Y}_{i+l},Y_{i}^{i+l-1}\right) \\
    & \quad -g_y\left(\tilde{Y}_{i+l},Y_{i}^{i+l-1}\right) \Big{|} \numberthis.
\end{align*}
Definition \ref{def:causal_function} is a sub-functions class of the continuous sequence to sequence functions class, and applies to the causal transformer. Thus, concludes that the causal transformer $g\in\GTFC^{(d_i,d_o,l,v)}$ also applies for the universal approximation theorem, for continuous vector-values functions. Moreover, for our case the output sequence is a scalar value, $\cU\subset \RR^{l\times d_i}, \cZ\subset \RR^{1\times d_o}$.
For given $\epsilon, M, l$ and $n$, denote $g_y^\star \in \GTFC^{(d_y, 1, l, v)}$ the causal transformer, such that the approximation error is uniformly bounded $\exp{[-M]}\times {\epsilon}/{4}$ for the final time prediction out from the model.
Finally, combining Theorem \ref{theorem:ua_transformers} we have
\begin{align*}
    &\left|\hDYT(\Dn) -\DYT \right|\\
    &\leq \left( 1+e^M\right)\frac{1}{n}\sum_{i=1}^{n}\Big{|} f_{y,l}^\star \left(\tilde{Y}_{i+l},Y_{i}^{i+l-1}\right) \\
    & \quad -g_y^\star\left(\tilde{Y}_{i+l},Y_{i}^{i+l-1}\right) \Big{|} + \frac{\epsilon}{4}\\
    &\leq \frac{\epsilon}{2}.
    \numberthis \label{eq:dy_end_proof}
\end{align*}
This concludes the proof of \eqref{eq:dy_dv_repres}. For \eqref{eq:dyx_dv_repres}, note that
\begin{align*}
    f_{y,l}^{\star}&:=\log \left( \frac{dP_{Y^l}}{d(P_{X^lY^{l-1}}\otimes \tilde{P}_{Y_l})}\right)\\
    &=\log p_{Y_l | X^lY^{l-1}}-\log \tilde{p}_{Y_l},\numberthis
\end{align*}
achieves the supremum. Following the same claims for $\hDYXT(\Dn)$,
\begin{equation}\label{eq:dxy_end_proof}
    \left|\hDYXT(\Dn) -\DYXT \right| \leq \frac{\epsilon}{2},
\end{equation}
where 
\begin{align*}
    &\hDYXT(\Dn)\\
    &:=\sup_{g_{xy}\in \GTF^{XY}}\frac{1}{n}\sum_{i=0}^{n-1}g_{xy}\left(Y_{i}^{i+l}, X_{i}^{i+l}\right)\\
    &\quad -\log\left( \frac{1}{n}\sum_{i=0}^{n-1}e^{g_{xy}\left(Y_{i}^{i+l-1},X_{i}^{i+l},\tilde{Y}_{i+l}\right)}\right).\numberthis
\end{align*}
With \eqref{eq:dxy_end_proof} and \eqref{eq:dy_end_proof} we end the proof. $\blacksquare$
\raggedbottom   
\subsection{Proof of lemma 2
}\label{subsec:te_dkl_proof}
\noindent
Recall
\begin{subequations}
    \begin{align*}
        &\DYT := \DKL \left(P_{Y_l|Y^{l-1}} \| \tilde{P}_{Y_l} \Big{|} P_{Y^{l-1}} \right),\numberthis\\
        &\DYXT := \DKL \left(P_{Y_l|Y^{l-1}X^l} \| \tilde{P}_{Y_l} \Big{|} P_{Y^{l-1}X^{l}} \right).\numberthis
    \end{align*}
\end{subequations}
By expanding the first term we obtain,
\begin{align*}
    &\DYT\\
    &= \EE_{P_{Y^{l-1}}} \left[\DKL(P_{Y_l|Y^{l-1}} \| P_{\tilde{Y}_l}) \right] \\
    &= \int_{\cY^{l-1}} \Bigg{[}\int_{\cY_l} \log \left( \frac{P(y_l|y^{l-1})}{\tilde{P}(y^{l})}\right) \\
    &  \quad P(y_l|y^{l-1})d(y_l) \Bigg{]} P(y^{l-1})d(y^{l-1})\\
    &= \int_{\cY^{l}} \log \left( \frac{P(y_l|y^{l-1})}{\tilde{P}(y_l)}\right) P(y^{l})d(y^{l})\\
    &= \int_{\cX^{l}\cY^{l}} \log \left( \frac{P(y_l|y^{l-1})}{\tilde{P}(y_l)}\right) P(x^{l}y^{l})d(x^{l}y^{l}),\numberthis
\end{align*}
and the second term,
\begin{align*}
    &\DYXT \\
    &= \EE_{P_{X^{l}Y^{l-1}}} \left[\DKL(P_{Y_l|X^{l}Y^{l-1}} \| P_{\tilde{Y}_l}) \right] \\
    &= \int_{\cX^{l}\cY^{l-1}}\Bigg{[}\int_{\cY_l} \log \left( \frac{P(y_l|x^{l}y^{l-1})}{\tilde{P}(y_l)}\right) \\
    &  \quad P(y_l|x^{l}y^{l-1})d(y_l) \Bigg] P(x^{l}y^{l-1})d(x^{l}y^{l-1})\\
    &= \int_{\cX^{l}\cY^{l}} \log \left( \frac{P(y_l|x^{l}y^{l-1}}{\tilde{P}(y_l)}\right) P(x^{l}y^{l})d(x^{l}y^{l}),\numberthis
\end{align*}
where $\tilde{P}_Y$ is a reference density function and is absolutely continuous on $\cY$.
Subtructing the two terms resulting,
\begin{align*}
    &\DYXT - \DYT \\
    &= \int_{\cX^{l}\cY^{l}} \log \left( \frac{P(y_l|x^{l}y^{l-1})}{P(y_l|y^{l-1}}\right) P(x^{l}y^{l})d(x^{l}y^{l}) \\
    &= h(Y_l | Y^{l-1}) - h(Y_l | X^{l} Y^{l-1}) \\
    &= \sTE(l). \quad \square \numberthis
\end{align*}

\subsection{Proof of lemma \ref{lemma:optimal_te}}\label{sec:proof_optimal_te}
\noindent
The optimal TE, $\sTE^\star(l)$, by non-finite memory process $\XX$, is given by 
\begin{equation}\label{eq:optimal_te_def}
    \sTE^\star(l):=\lim_{n\to \infty} \sup_{P_{X^l_{-n}}} \sTE(l).
\end{equation}
For any $n>0$ we have,
\begin{align*}\label{eq:optimal_te}
    &\sup_{P_{X^l_{-n}}} \sTE(l) \\
    &=\sup_{P_{X^0_{-n}}P_{X^l | X^0_{-n}}} \sI\left( X^l;Y_l | Y^{l-1} \right)\\
    &=\sup_{P_{X^0_{-n}}P_{X^l | X^0_{-n}}}
        \int_{\cX^{l}_{-n}\cY^{l}} P(x^0_{-n}) P(x^l,y^l | x^0_{-n}) \\
        & \quad \underbrace{\sI \left(  X^l;Y_l | Y^{l-1} \right)}_{\text{independent of } x^{0}_{-n}} d(x^l x^{0}_{-n} y^l) \\
    &\stackrel{(a)}{=}\sup_{P_{X^l}}
        \int_{\cX^{l}\cY^{l}} P(x^l,y^l)\sI \left(  X^l;Y_l | Y^{l-1} \right) d(x^l y^l)  \\
    &=\sup_{P_{X^l}} \sTE(l), \numberthis
\end{align*}
where (a) follows from the fact that the conditional MI does not depend on $x^0_{-n}$, thus we can eliminate the integral of $\cX_{-n}^{0}$ that does not affect the conditional MI, as for the supremum. 
Since \eqref{eq:optimal_te} is true for any $n\in\NN$, with \eqref{eq:optimal_te_def} we get,
\begin{equation}\label{eq:optimal_te_fin}
    \sTE^\star(l)= \sup_{P_{X^l}} \sTE(l). \quad \square
\end{equation}

\subsection{List of Symbols}\label{sec:appendix_symbols}
\noindent {\raggedright The following table summarizes all primary symbols for quick reference.}
\begin{center}
\vspace{0.5em}
\begingroup
\setlength{\tabcolsep}{4pt}
\setlength{\extrarowheight}{2pt}
\footnotesize
\begin{supertabular}{@{}lp{0.65\linewidth}@{}}
    \toprule
    Symbol & Definition \\
    \midrule
    $\RR, \ZZ, \NN$              
    & Sets of real numbers, integers, and natural numbers, respectively.\\
    $\cX$                         
    & Subset of $\RR^d$, the observation space.\\
    $\XX$                         
    & Stochastic process $(X_t)_{t\in\ZZ}$.\\
    $\PP$                         
    & Probability measure on $(\Omega,\cF)$.\\
    $\cP(\cX)$                   
    & Set of Borel probability measures on $\cX$.\\
    $\lebmeas(\cX)$              
    & Subset of $\cP(\cX)$ absolutely continuous w.r.t.\ Lebesgue measure.\\
    $\EE$                         
    & Expectation operator under $\PP$.\\
    $\sh_{\mathsf{CE}}(P,Q)$     
    & Cross‐entropy between $P$ and $Q$.\\
    $\sh(X)$                      
    & Differential entropy of $X\sim P$, i.e.\ $\sh_{\mathsf{CE}}(P,P)$.\\
    $\DKL(P||Q)$                 
    & Kullback–Leibler divergence between $P$ and $Q$.\\
    $\sI(X;Y)$                    
    & Mutual information between random variables $X,Y$.\\
    $\sTE(l)$          
    & Transfer entropy from process $X$ to $Y$ with memory parameter $l$.\\
    $\sI(\XX\to\YY)$           
    & Directed information from $\XX$ to $\YY$.\\
    $W$                        
    & Weight matrices in neural network layers.\\
    $b$                     
    & Bias vectors in neural network layers.\\
    $\alpha_{ij}$                 
    & Attention weight from position $i$ to $j$.\\
    $Q,K,V$                       
    & Query, key, and value matrices in self‐attention.\\
    $\attn(\cdot)$               
    & Dot‐product attention function (Definition~\ref{def:attention_dotproduct}).\\
    $\mathsf{softmax}(\cdot)$    
    & Softmax activation applied columnwise.\\
    $\Xpe$                       
    & Positional‐encoded for input $X$.\\
    $M_{[i,j]}$ 
    & Causal mask entry: can be 1 or $-\infty$ to include or ignore positions in the softmax operation.\\
    $\cR_{(i,j)}$
    & Attention coefficient weight from query at time $i$ to key at time $j$. \\
    $\GTF^{(d_i,d_o,l,v)}$        
    & Class of transformers with input dim $d_i$, output dim $d_o$, length $l$, and width $v$ (Definition~\ref{def:transformer_func_class}).\\
    $\GTFC^{(d_x,1,l,v_y)}$ 
    & Class of causal transformer architectures for single-process and joint-process estimation (Definition~\ref{def:transformer_func_class}). \\
    \shortstack{$\DYT$\\$\DYXT$} 
    & Variational KL objectives (\eqref{eq:dy_dyxt_def}). \\
    $\tene(\Dn;l)$               
    & TREET estimator objective for sample set $\Dn$ and memory $l$.\\
    \shortstack{$\hDYT$\\ $\hDYXT$}
    & Empirical estimators of KL objectives (\eqref{eq:tene_objective_estimator}). \\
    $C_{\ff}$  
    & Feedforward channel capacity. \\
    $C_{\fb}$  
    & Feedback channel capacity. \\
    \bottomrule
\end{supertabular}
\captionof{table}{Comprehensive list of symbols used throughout the manuscript.}\label{tab:appendix_symbols}
\endgroup
\end{center}

\subsection{Implementation Parameters}\label{sec:appendix_implementation}
\noindent
For the TE benchmark, channel capacity estimation model, and density estimation, we trained the optimization procedure with a limit of 200 epochs. We utilized a batch size of 1024, a learning rate of $8 \times 10^{-3}$, and the Adam optimizer \cite{kingma2014adam}. The transformer architecture comprises one attention layer followed by a FF layer. The attention mechanism is implemented with a single head, having a dimension of 32 neurons. The dimension of the FF layer is 64, featuring ELU activation \cite{clevert2015fast}. Typically the order of TE $l$, was set to 30. The inputs sequence length was set to $l+30$ to calculate 30 series in parallel. Each epoch generates 100K samples of $\XX, \YY$ with corresponding random noise (default is uniform). For the cases of optimization, the NDG also employs a transformer structure with one attention layer and one FF layer, maintaining the same dimensional specifications. The learning rate for the NDG is set to $8 \times 10^{-4}$. It is trained at a rate of once every four epochs.
For the Apnea dataset feature analysis, the learning rate was adjusted to $1 \times 10^{-4}$. Additionally, we configured the model to use 1 head with a dimension of 16 for the attention mechanism, and the FF layer dimension was set to 32, with ELU activation.
All experiments were conducted on \textit{Nvidia RTX 3090 and RTX Ada 6000 GPUs}. 


\bibliographystyle{unsrt}
\bibliography{ref.bib}

\end{document}